\documentclass[aps,preprintnumbers,superscriptaddress,onecolumn,amsmath,amssymb,floatfix,pra]{revtex4-1}

\pdfoutput=1
\usepackage{graphicx}
\usepackage[usenames,dvipsnames]{xcolor}
\usepackage{braket}
\usepackage[colorlinks,bookmarks=false,citecolor=violet,linkcolor=Red,urlcolor=blue]{hyperref}
\usepackage[export]{adjustbox}
\usepackage{todonotes}

\usepackage{mathtools}

\usepackage{tikz}

\usepackage{braket}

\newcommand\ac[1]{a_{\mathcal{C}}(#1)}

\newcommand{\ci}{\mathrm{i}\mkern1mu}
\newcommand{\re}{\textrm{Re}\,}
\newcommand{\im}{\textrm{Im}\,}
\newcommand{\ve}{\varepsilon}
\newcommand{\tve}{\tilde{\varepsilon}}
\begin{document}

\title{Non-probabilistic fermionic limit shapes}
\author{Saverio Bocini}
\affiliation{Univ Lyon, CNRS, Universit\'e Claude Bernard Lyon 1, UMR5208, Institut Camille Jordan, F-69622 Villeurbanne, France}
\affiliation{Dipartimento di Fisica, Universit\'a di Firenze Via G. Sansone 1, 50019 Sesto Fiorentino - Firenze, Italy}

\author{Jean-Marie St\'ephan}
\affiliation{Univ Lyon, CNRS, Universit\'e Claude Bernard Lyon 1, UMR5208, Institut Camille Jordan, F-69622 Villeurbanne, France}

\begin{abstract}
 We study a translational invariant free fermions model in imaginary time, with nearest neighbor and next-nearest neighbor hopping terms, for a class of inhomogeneous boundary conditions. This model is known to give rise to limit shapes and arctic curves, in the absence of the next-nearest neighbor perturbation. The perturbation considered turns out to not be always positive, that is, the corresponding statistical mechanical model does not always have positive Boltzmann weights. We investigate how the density profile is affected by this nonpositive perturbation. We find that in some regions, the effects of the negative signs are suppressed, and renormalize to zero. However, depending on boundary conditions, new ``crazy regions'' emerge, in which minus signs proliferate, and the density of fermions is not in $[0,1]$ anymore. We provide a simple intuition for such behavior, and compute exactly the density profile both on the lattice and in the scaling limit. 
\end{abstract}

\maketitle

\newpage
\section{Introduction}

Most models in two dimensional statistical mechanics have bulk properties that are homogeneous and do not depend on boundary conditions. For example free energy, or local magnetization in spin models, typically show this behavior. Such properties should not be taken for granted however. There are models, in particular with underlying (continuous) symmetries, which violate this principle. These show strong dependence on boundary conditions, and a nontrivial density or magnetization profile. It may even happen that some regions are frozen: they have zero entropy in the thermodynamic limit, or, said differently, they become deterministic. The curve separating the frozen region from the non trivial fluctuating region is called ``arctic curve''. Its determination is in general challenging. The nontrivial (integrated) density profile is often referred to as ``limit shape'' in the literature.

The  emergence of such nontrivial limit shapes has attracted a lot of interest, in physics, mathematics, and computer science, with perhaps the most famous instance provided by the arctic circle theorem \cite{ArcticCircle}, which concerns dimer coverings of an Aztec diamond \cite{Elkies1991,Elkies1992}. Another example of an arctic circle is shown in figure \ref{fig:densityprofiles}(a). Older instances can also be found in the context of Young diagrams \cite{kerov_vershik} and crystal shapes \cite{Pokrovsky_Talapov}. Over the years, many connections to stochastic processes \cite{Spohn_prl}, random matrices \cite{Johansson2000,johansson2005}, representation theory \cite{2012BorodinGorin}, variational \cite{Nienhuis_1984,variationaldimers,kenyon2007} or hydrodynamic \cite{Abanov_hydro} ideas, and even algebraic geometry \cite{Amoeba} have been investigated.

Limit shapes in two dimensions can also be understood as one-dimensional quantum systems evolving in imaginary time \cite{Reshetikhin,Allegra_2016,Stephan_ebc}, through the transfer matrix formalism \cite{Onsager_1944}. In fact, essentially all models that have been solved analytically can be seen through this prism, with an underlying Hamiltonian which turns out to be integrable.  While the most heavily studied setups map to free fermions, progress on understanding the effects of interactions has been slow but steady, with a pace that has been increasing in recent years  \cite{Korepin1982,Izergin1987,ColomoPronko_arctic,ColomoPronkoZinnJustin,TangentMethod,borodin2016,Reshetikhin,Granet_2019,deGierKenyon,Aggarwal_iceproof,Debin2020,KRS}. This problem is also related to ground state properties of well-known spin chains or quantum gases in inhomogeneous trapping potentials (e.g. \cite{BrunDubail2018}), treated through the local density approximation \cite{cold_review}.

Another motivation for studying limit shapes lies in the relation --through the Wick rotation-- to real-time quantum quench protocols. It may happen for certain complicated observables that the real time problem is intractable while the corresponding imaginary time treatment is much simpler. This occurs for example when studying the entanglement entropy of quantum systems that can be tackled by conformal field theory methods \cite{Calabrese_2016}. To make progress, one first assumes time is imaginary, compute those observables within euclidean path integral combined with a replica trick, and then simply ``recall'' that time is real. While there is of course no mathematical justification for this method in general, it has led to highly non trivial conjectures, especially regarding the growth of entanglement. These conjectures are then typically checked numerically to high precision  \cite{calabrese2007entanglement,calabrese2005time,calabrese2006time,stephan2011local,Collura_2013,dubail2017conformal}, leaving no doubt that the final analytical prediction is correct.

Of course, not every quantum Hamiltonian maps to a statistical model with positive Boltzmann weights, and this well-known observation motivates the present study. Such types of non positive models are often encountered rather directly when simulating quantum systems, using Monte Carlo methods \cite{becca_sorella_2017}. This is often dubbed ``sign problem'', as the lack of positivity of the statistical mechanical interpretation makes simulations extremely slow to converge, similar to the difficulties encountered while numerically integrating highly oscillatory functions. Statistical models with non positive weights are also interesting in their own rights, in particular they often appear when trying to reformulate a non-local but positive model (typically a loop model) as a local one. In that case the price to pay is loss of positivity. This has important consequences at the critical point, where complicated and still poorly understood logarithmic CFTs emerge \cite{Gainutdinov_2013}.

In this paper, we study limit shapes that are not necessarily positive. Our starting point is a one-dimensional free fermion chain with nearest neighbor hoppings, known also as the XX chain (after Jordan-Wigner transformation). We look at it in imaginary time, with a simple class of boundary conditions. This is well known to give rise to limit shapes, positive and related to the polynuclear growth (PNG) stochastic model \cite{Spohn_prl,PraehoferSpohn2002}. We then perturb the Hamiltonian with a next-nearest neighbor fermion hopping, which can be shown breaks positivity except at special points which we also study. This is arguably one of the simplest possible example of a non positive model, which is nontrivial and can be solved. Our focus is on the resulting density profile, which we compute exactly, both on the lattice and in the thermodynamic limit.  

\newpage
\section{A fermion model with next nearest neighbor hoppings}
\subsection{Setup}
The problem we are looking at is a system of fermions governed by the following tight-binding Hamiltonian:
\begin{equation}\label{eq:themodel}
 H=\frac{1}{2}\sum_{x\in \mathbb{Z}}\left[c_{x+1}^\dag c_x+c_x^\dag c_{x+1}+\alpha\left(c_{x+2}^\dag c_x+c_x^\dag c_{x+2}\right)\right]
\end{equation}
where the $c_x, c_x^\dag$ are fermionic operators with anticommutation relations $\{c_x,c_{x'}^\dag\}=\delta_{xx'}$, $\{c_x,c_{x'}\}=\{c_x^\dag,c_{x'}^\dag\}=0$. Lattice sites are on the infinite discrete line $\mathbb{Z}$. We assume $\alpha\geq 0$ \emph{throughout the whole paper}. In momentum space, the Hamiltonian reads
\begin{equation}
 H=\int_{-\pi}^{\pi} \frac{dk}{2\pi}\varepsilon(k) c^\dag(k)c(k)\quad,\qquad  c^\dag(k)=\sum_{x\in \mathbb{Z}}e^{\ci k x}c_x^\dag,
\end{equation}
with dispersion
\begin{equation}
 \varepsilon(k)=\cos k +\alpha \cos 2k.
\end{equation}
Such deformations have been considered in different contexts, see e.g \cite{Zvonarev2003}.
We look at the model in imaginary time, following \cite{Allegra_2016}. The expectation values of local observables are defined as
\begin{equation}\label{eq:Odef}
 \braket{O_x(y)}=\frac{\braket{\psi_n|e^{(R-y)H}O_x e^{(R+y)H}|\psi_n}}{\braket{\psi_n|e^{2RH}|\psi_n}}.
\end{equation}
For $\alpha=0$ the interpretation as a statistical model in a slab geometry is well-known \cite{PraehoferSpohn2002,Allegra_2016}, see figure \ref{fig:densityprofiles}(a). Here $2R>0$ is the width of the slab, and $y\in [-R,R]$ is a vertical coordinate inside the slab. The boundary conditions are imposed by the initial and final state, which we take to be identical. Our focus is on the following particular class of real-space product states
\begin{align}
 \ket{\psi_n}&=\prod_{x\leq 0} c_{nx}^\dag \ket{0}\\
 &=\ldots  c_{-3n}^\dag c_{-2n}^\dag c_{-n}^\dag c_0^\dag \ket{0},
\end{align}
where the fermions are ordered from left to right, $\ket{0}$ is the vacuum, and $n$ is an integer $\geq 1$. In $\ket{\psi_n}$ all sites to the right of the origin are empty, while on the left there is a fermion every $n$ site. This inhomogeneous initial state has average density $d=1/n$ to the left, $0$ to the right, which leads to nontrivial behavior. 
The justification for studying those in particular is that they will be sufficient for our purposes, while remaining technically reasonably simple. 
Other more complicated states are possible too, and would lead to similar conclusions.

To lift possible ambiguities \footnote{For example, the partition function on the denominator of (\ref{eq:Odef}) is infinite for $n\geq 2$, but the ratio is finite.} or to perform numerical simulations, it is convenient to consider a finite volume version of $H$ and $\ket{\psi_n}$ to sites $\{-L,-L+1,\ldots,L-1,L\}$ with open boundary conditions. Then one computes expectation values $\braket{O_x(y)}_L$, in finite volume, with (\ref{eq:Odef}) defined as the $L\to \infty$ limit. In the following, we focus most of our attention on the local ``density'' of fermions $O_x=c_x^\dag c_x=n_x$.
\subsection{Negative states}
As will be shown below, the model given by (\ref{eq:themodel}),(\ref{eq:Odef}) is not necessarily positive (the underlying statistical mechanical model would not be guaranteed to have positive Boltzmann weights) for $\alpha>0$.  
Before proceeding let us already present some numerical results, which give an idea of the effect of the second neighbor term for large values of $R$.
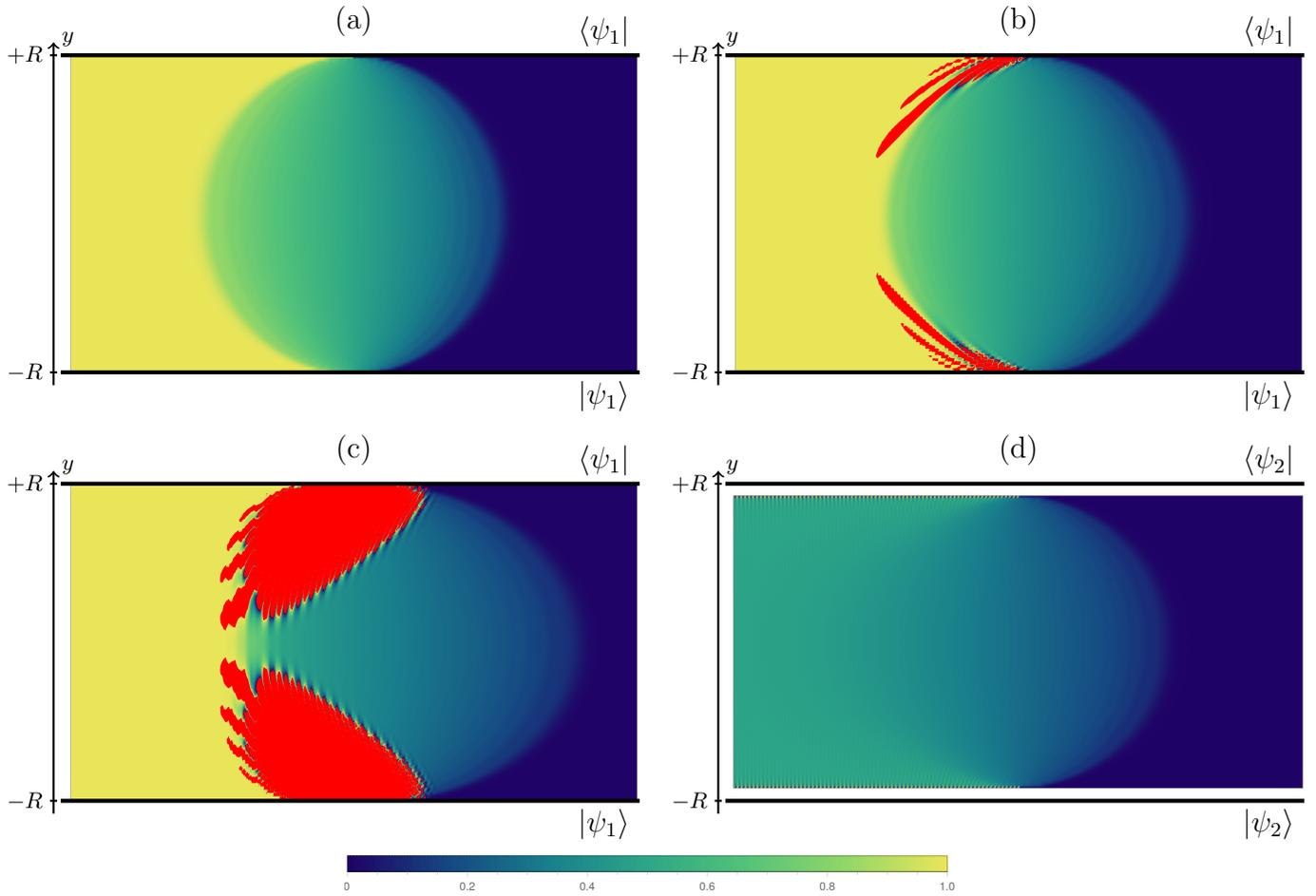
\begin{figure}[htbp]
\begin{tikzpicture}
\node at (0,0) {\includegraphics[width=0.46\textwidth]{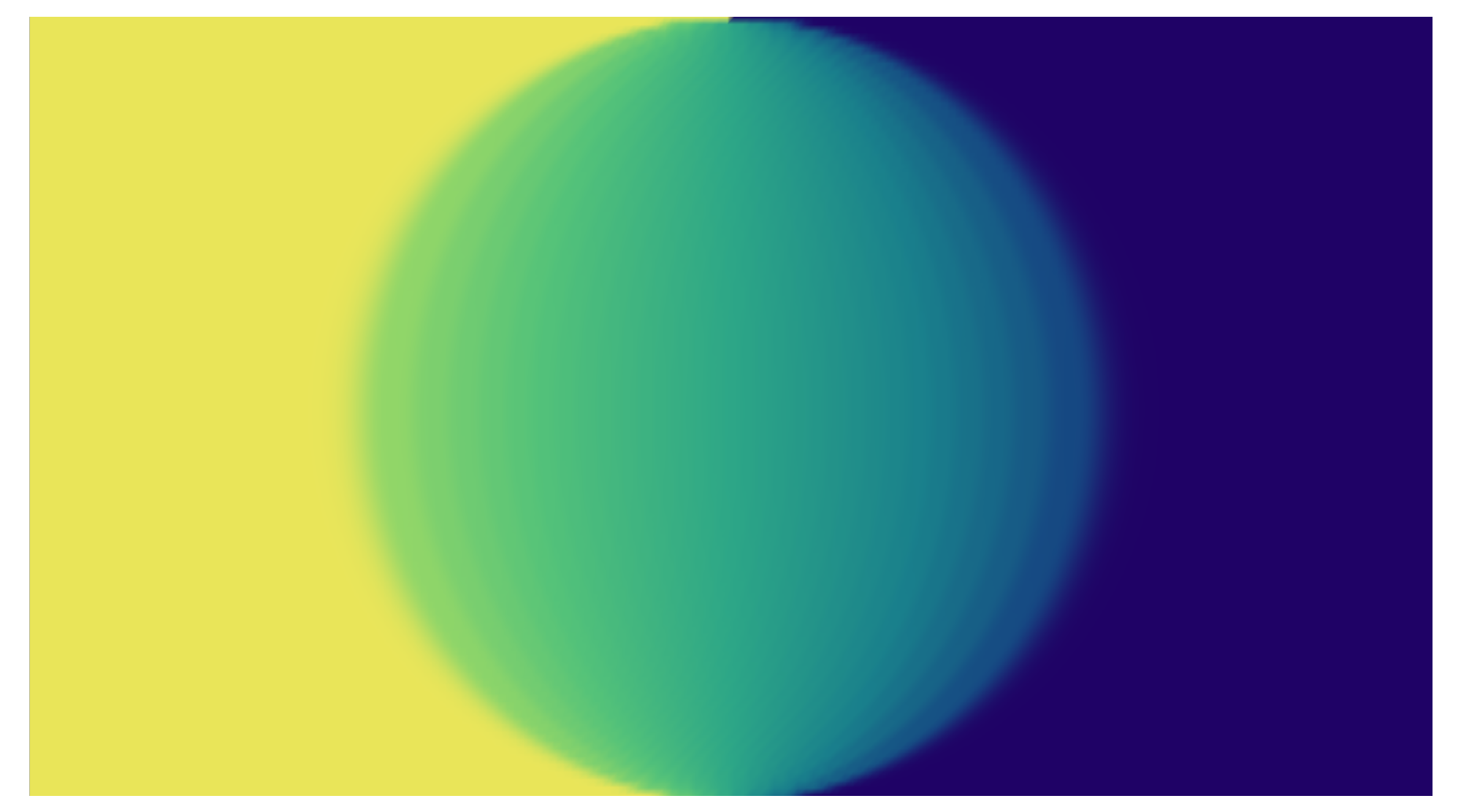}};
\node at (9.3,0) {\includegraphics[width=0.46\textwidth]{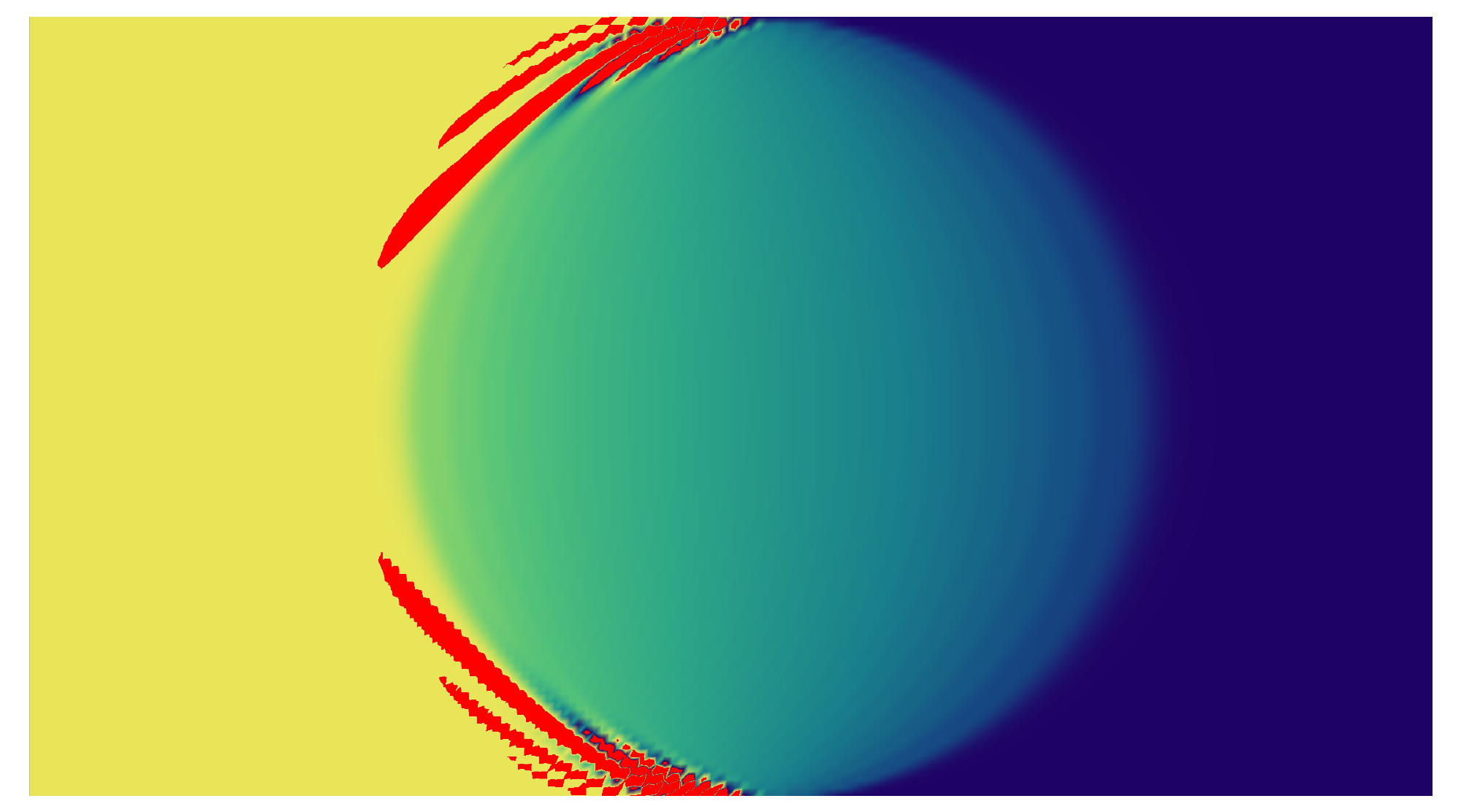}};
\node at (0,-6) {\includegraphics[width=0.46\textwidth]{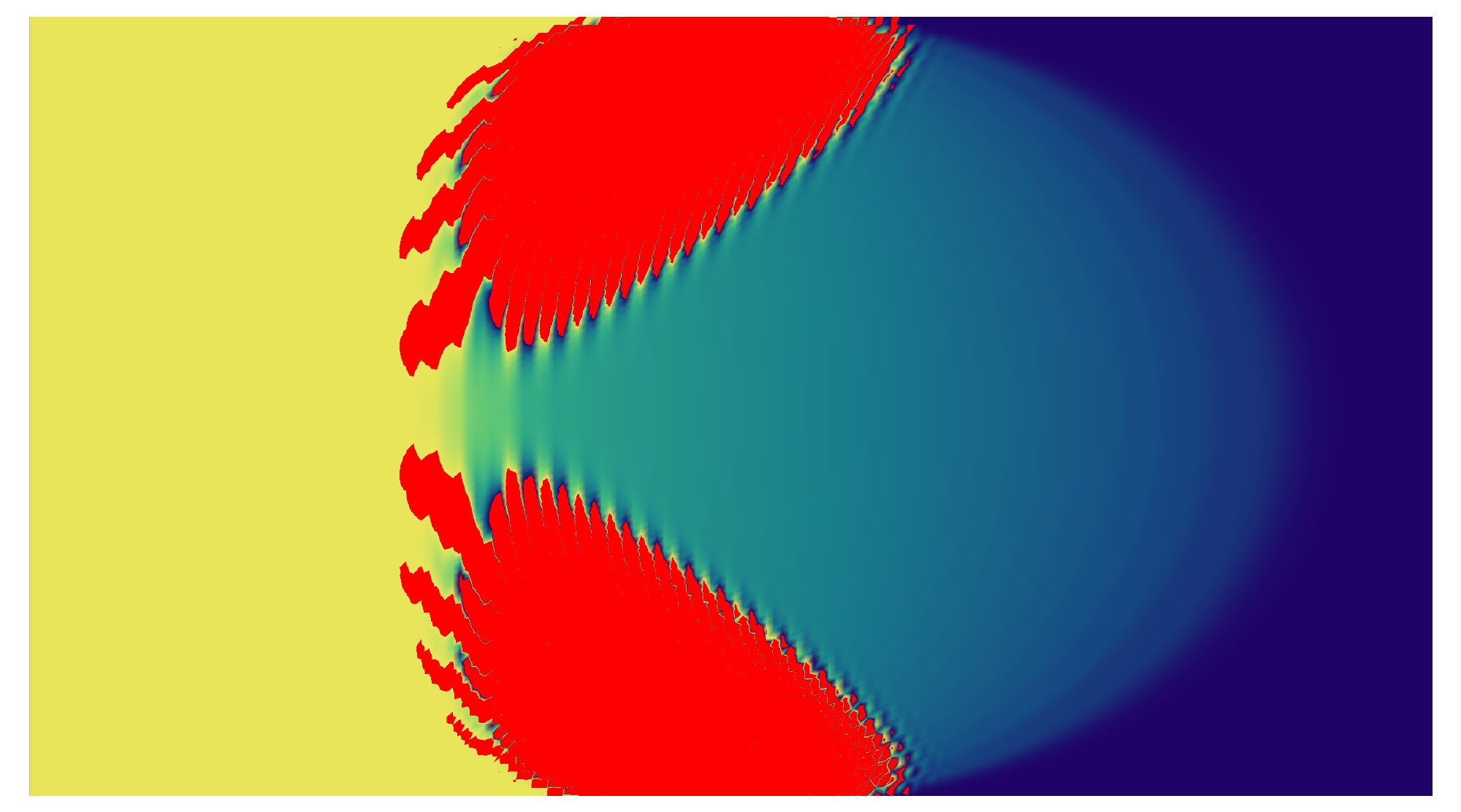}};
\node at (9.3,-6) {\includegraphics[width=0.445\textwidth]{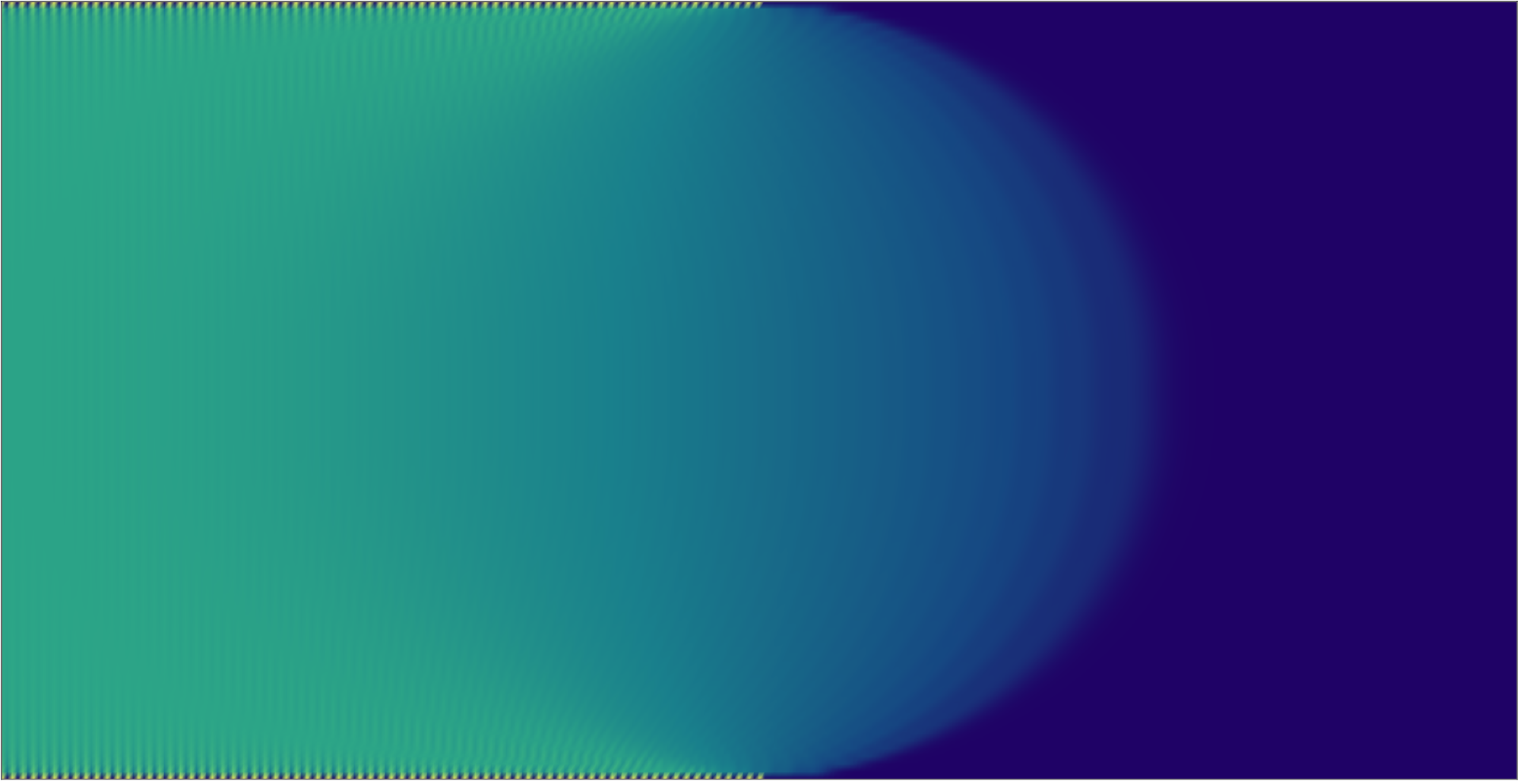}};
\draw[ultra thick] (-4.1,-2.22) -- (4.,-2.22);
\draw[ultra thick] (-4.1,2.22) -- (4.,2.22);
\draw (3.5,-2.55) node {\large{$\ket{\psi_1}$}};
\draw (3.5,2.55) node {\large{$\bra{\psi_1}$}};
\draw[thick,->] (-4.2,-2.4) -- (-4.2,2.4);
\draw[thick] (-4.25,-2.22) -- (-4.15,-2.22);
\draw[thick] (-4.25,2.22) -- (-4.15,2.22);
\draw (-4.,2.45) node {$y$};
\draw (-4.6,2.22) node {$+R$};
\draw (-4.6,-2.22) node {$-R$};
\draw (0,2.7) node {\large{(a)}};
\begin{scope}[xshift=9.3cm]
 \draw[ultra thick] (-4.1,-2.22) -- (4.,-2.22);
\draw[ultra thick] (-4.1,2.22) -- (4.,2.22);
\draw (3.5,-2.55) node {\large{$\ket{\psi_1}$}};
\draw (3.5,2.55) node {\large{$\bra{\psi_1}$}};
\draw[thick,->] (-4.2,-2.4) -- (-4.2,2.4);
\draw[thick] (-4.25,-2.22) -- (-4.15,-2.22);
\draw[thick] (-4.25,2.22) -- (-4.15,2.22);
\draw (-4.,2.45) node {$y$};
\draw (-4.6,2.22) node {$+R$};
\draw (-4.6,-2.22) node {$-R$};
\draw (0,2.7) node {\large{(b)}};
\end{scope}
\begin{scope}[yshift=-6cm]
 \draw[ultra thick] (-4.1,-2.22) -- (4.,-2.22);
\draw[ultra thick] (-4.1,2.22) -- (4.,2.22);
\draw (3.5,-2.55) node {\large{$\ket{\psi_1}$}};
\draw (3.5,2.55) node {\large{$\bra{\psi_1}$}};
\draw[thick,->] (-4.2,-2.4) -- (-4.2,2.4);
\draw[thick] (-4.25,-2.22) -- (-4.15,-2.22);
\draw[thick] (-4.25,2.22) -- (-4.15,2.22);
\draw (-4.,2.45) node {$y$};
\draw (-4.6,2.22) node {$+R$};
\draw (-4.6,-2.22) node {$-R$};
\draw (0,2.7) node {\large{(c)}};
\end{scope}
\begin{scope}[xshift=9.3cm,yshift=-6cm]
 \draw[ultra thick] (-4.1,-2.22) -- (4.,-2.22);
\draw[ultra thick] (-4.1,2.22) -- (4.,2.22);
\draw (3.5,-2.55) node {\large{$\ket{\psi_2}$}};
\draw (3.5,2.55) node {\large{$\bra{\psi_2}$}};
\draw[thick,->] (-4.2,-2.4) -- (-4.2,2.4);
\draw[thick] (-4.25,-2.22) -- (-4.15,-2.22);
\draw[thick] (-4.25,2.22) -- (-4.15,2.22);
\draw (-4.,2.45) node {$y$};
\draw (-4.6,2.22) node {$+R$};
\draw (-4.6,-2.22) node {$-R$};
\draw (0,2.7) node {\large{(d)}};
\end{scope}
\end{tikzpicture}
\vspace{0.2cm}
\includegraphics[width=9cm]{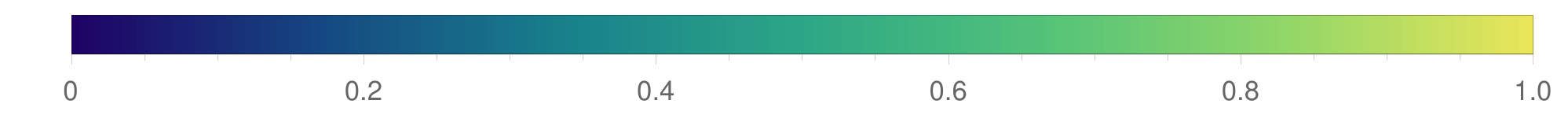}
 \caption{Numerical density profiles for $R=50$. (a): arctic circle, $\alpha=0$, $\ket{\psi_1}$. (b): $\alpha=1/15$, $\ket{\psi_1}$. (c): $\alpha=1/4$, $\ket{\psi_1}$. (d): $\alpha=1/2$, $\ket{\psi_2}$. The color code is shown at the very bottom: yellow means density one, while blue means density $0$. Intermediate colors interpolate. The neon red lies outside this scale, and corresponds to regions which have an ill-defined density, not in $[0,1]$ anymore. We call them crazy regions.}
 \label{fig:densityprofiles}
\end{figure}

Figure \ref{fig:densityprofiles}(a) shows the density profile for the (probabilistic, $\alpha=0$) XX chain, which is an instance of the arctic circle \cite{Spohn_prl,Allegra_2016}. 
For the other subfigures, the density seems to be well defined for large $R$, but only in some regions in the $(x,y)$ plane. However, new ``crazy regions'' (in red) can be seen in figure \ref{fig:densityprofiles}(b,c), for $\ket{\psi_1}$ and $\alpha=1/15,1/4$, in which the density is not in $[0,1]$ anymore. Worse, we observe that it oscillates wildly, with an amplitude that appears to grow exponentially fast with $R$. The size of the crazy region seems also to grow when increasing $\alpha$.
 Therefore, the effect seems either (exponentially) suppressed, or completely destabilize the probabilistic nature of the model at $\alpha=0$, leading to seemingly nonsensical density. 
Finally, the density profile for $\alpha=1/2$ and $\ket{\psi_2}$ is shown in figure \ref{fig:densityprofiles}(d), where crazy regions are nowhere to be seen, an observation which appears to hold for any $\alpha\geq 0$. 

Even though all terms in (\ref{eq:themodel}) appear with a plus sign, the model is not positive for $\alpha> 0$ and $y\neq 0$, as can be shown by the following argument (we set $n=1$ and consider a finite chain for simplicity). The time-evolved state $\ket{\psi_1(\tau)}=e^{\tau H}\ket{\psi_1}$ can be written formally by inserting a complete basis of real-space states $\ket{\mathcal{C}}=c_{i_1}^\dag \ldots c_{i_p}^\dag \ket{0}$ where the fermions are ordered $i_1<\ldots<i_p$. In the following we also find it convenient to use the notation 
\begin{equation}
 c_{i_1}^\dag \ldots c_{i_p}^\dag \ket{0}=\ket{\ldots 0\underbrace{1}_{i_1}0\ldots 0 \underbrace{1}_{i_p}0\ldots}
\end{equation}
where a $1$ denote the presence of a fermion, a $0$ the absence of a fermion. Coming back to the time-evolved state
\begin{equation}
 \ket{\psi_1(\tau)}=\sum_{\mathcal{C}} a_\mathcal{C}(\tau)\ket{\mathcal{C}}\qquad,\qquad a_\mathcal{C}(\tau)=\braket{\mathcal{C}|e^{\tau H}|\psi_1},
\end{equation}
 each coefficient has a power series representation  
\begin{align}
 a_\mathcal{C}(\tau)&=\sum_{m=0}^\infty \frac{\tau^m}{m!} \braket{\mathcal{C}|H^m| \psi_1}.
\end{align}
When $\alpha=0$ there are only nearest-neighbor hoppings, so any number of applications of $H$ on $\ket{\psi_1}$ --or any other real space product state $\ket{\psi}$-- cannot change the order of the fermions. Hence one gets a linear combination of basis states with positive coefficients $a_\mathcal{C}(\tau)$. For $\alpha>0$ however, we have for example
\begin{align}
2H\ket{\psi_1}&=2H\ket{\ldots 11111\,00000\ldots}\\&=\label{eq:someminus}
\ket{\ldots 11110\,10000\ldots}+\alpha \ket{\ldots 11110\,01000\ldots}-\alpha \ket{\ldots 11101\,10000\ldots}, 
\end{align}
 so this next nearest neighbor perturbation produces both positive and negative terms. The last one is due to the fact that the second rightmost fermion hop around the rightmost fermion, producing a minus sign due to anticommutation relations. For sufficiently small $\tau$ this ensures that some $a_\mathcal{C}(\tau)$ are negative. In fact, very few types of dispersions give nonnegative $\ac{\tau}$ for arbitrary real-space product initial states. Those dispersions can be classified thanks to the Edrei-Thoma theorem \cite{Aissen1952,Edrei,Thoma} on totally nonnegative Toeplitz matrices. The relation to total nonnegativity has been emphasized e.g. in the context of cluster algebras \cite{DiFrancesco} and determinantal point processes \cite{BerggrenDuits} related to our model. We recall how this applies to free fermions in appendix \ref{app:totalpos}, where we also discuss further several related notions of positivity in fermionic models. 
 
The quantum mechanical average (\ref{eq:Odef}) corresponds to a ``distribution'' for the configurations along a horizontal line 
\begin{equation}\label{eq:nonproba_dist}
 \mathbb{P}(\mathcal{C})=\frac{a_\mathcal{C}(R-y)a_\mathcal{C}(R+y)}{\sum_{\mathcal{C}} a_\mathcal{C}(R-y)a_\mathcal{C}(R+y)},
\end{equation}
which is not guaranteed to be positive, unless $y=0$. Exploiting the fact that $\ac{\tau}$ can be computed exactly using Wick's theorem, we checked that $\mathbb{P}(\ket{\ldots1101\,1000\ldots})$ is negative for some values of $y$, as suggested by (\ref{eq:someminus}). Determining which configurations are negative, which are not, for which initial state, and whether they are suppressed or not in the thermodynamic limit is a non trivial question. In this paper, we narrow down the problem by  exploring their influence on the density profile
\begin{equation}
 \braket{n_x(y)}=\frac{\sum_{\mathcal{C}}'a_\mathcal{C}(R-y)a_\mathcal{C}(R+y)}{\sum_{\mathcal{C}}a_\mathcal{C}(R-y)a_\mathcal{C}(R+y)},
\end{equation}
where $\sum_{\mathcal{C}}'$ is the sum restricted to configurations $\mathcal{C}$ with a fermion at position $x$, $\ket{\mathcal{C}}=\ket{\ldots \underbrace{1}_x\ldots}$. 
We also present new results for the probabilistic point $y=0$. We focus on the following scaling limit $R\to \infty$, where the ratios $X=\frac{x}{R}\in \mathbb{R}$, $Y=\frac{y}{R} \in (-1,1)$ are kept fixed, in which case one expects density to become a continuous function of both $X$ and $Y$.

Let us come back to the results shown in figure \ref{fig:densityprofiles}.
A qualitative understanding can be achieved by noticing that negative signs occur when applying $H$ to states where a region with high density is connected to a region with lower density, since one needs two neighboring fermions for one to hop around the other, and empty space for the jump to be allowed. This is indeed what can be observed in the figure, where the crazy region connects to the domain wall, with a rough tendency to go west  where density is higher. On the other hand, the minus signs should be much less relevant when fermion density is much lower, on the east. This dilution argument makes it also plausible that density is always well-behaved for $\ket{\psi_{n\geq 2}}$ in the scaling limit, consistent with what is observed numerically. The model is however still not positive, at least for finite $L$, with the first negative terms occurring (say for $L\geq 32$) at order $\tau^6$, $\tau^7$ or $\tau^8$, depending on $\alpha$.

\subsection{An exact formula}
In this paper, we investigate this type of questions by performing exact calculations. 
Our main angle of attack is the following contour integral formula for the general propagator
\begin{equation}\label{eq:main_exact}
\frac{\braket{\psi_n|e^{(R-y') H}c_{x}^\dag  e^{(y'-y) H} c_{x'}e^{(R+y) H}|\psi_n}}{\braket{\psi_n|e^{2RH}|\psi_n}}=\int_{-\pi}^{\pi}\frac{dk}{2\pi}\int_{-\pi+\ci \eta}^{\pi+\ci \eta} \frac{dq}{2\pi} \frac{e^{\Phi_n(k,x,y)-\Phi_n(q,x',y')}e^{\Omega_n(k)+\Omega_n(q)}}{1-e^{-\ci n(k-q)}}
\end{equation}
where $\eta>0$,
\begin{equation}
 \Phi_n(k,x,y)=-\ci k x- y \varepsilon(k)+\ci R \tilde{\varepsilon}_n(nk),
\end{equation}
and
\begin{equation}
 \Omega_n(k)=R\left[\varepsilon(k)-\varepsilon_n(nk)\right].
\end{equation}
We have also introduced an effective dispersion
\begin{equation}\label{eq:effective_disp}
 \varepsilon_n(k)=\frac{1}{2R}\log\left(\frac{1}{n}\sum_{p=0}^{n-1}e^{2R\varepsilon(\frac{k+2p\pi}{n})}\right),
\end{equation}
with $\tilde{\varepsilon}_n$ its (periodic) Hilbert transform \footnote{For a periodic function of the form $f(k)=\sum_{p\geq 1}a_p \cos (pk)$, the periodic Hilbert transform is simply given by $\tilde{f}(k)=\sum_{p\geq 1}a_p \sin (pk)$}. Note $\varepsilon_1(k)=\varepsilon(k)$, so $\Omega_1(k)=0$. Note also $\varepsilon_n(k+2\pi)=\varepsilon_n(k)$.

The derivation of formula (\ref{eq:main_exact}) is presented in appendix \ref{app:derivation}. It relies solely on Wick's theorem, as well as linear algebra and Fourier analysis techniques which are presented in appendix \ref{app:toeplitz}. This result bears some similarity to a general result of Okounkov and Reshetikhin for Schur processes \cite{okounkovreshetikhin,okounkovreshetikhin2}, as well as others on determinantal point processes which can be found in the mathematical literature, see e.g. \cite{Johansson_randommatrices}. The case $n=1,\alpha=0$ can already be found in \cite{Spohn_prl}. For $n\geq 2$, the appearance of the extra effective dispersion (\ref{eq:effective_disp}) can be traced back to the fact that the $j$-th Fourier coefficient of $e^{2R\varepsilon_n(k)}$ is the $nj$-th Fourier coefficient of $e^{2R\varepsilon(k)}$.


We study the asymptotics of (\ref{eq:main_exact}) using saddle point techniques, first focusing on the case $n=1$ (section \ref{sec:dw}), before attacking $n\geq 2$ ( section \ref{sec:nw}).

\section{The domain wall boundary conditions}
\label{sec:dw}
For the domain wall case ($n=1$), the main formula (\ref{eq:main_exact}) can be rewritten in a simpler way. Indeed, we have $\varepsilon_1(k)=\varepsilon(k)$, and the Hilbert transform is simply obtained by replacing cosines with sines:
\begin{equation}
 \tve(k)=\sin k+\alpha \sin 2k.
\end{equation}
We also introduce the density
\begin{equation}
 \rho(X,Y)=\frac{\braket{\psi_n|e^{R(1-Y) H}c_{RX}^\dag c_{RX}e^{R(1+Y) H}|\psi_n}}{\braket{\psi_n|e^{2RH}|\psi_n}},
\end{equation}
in terms of the rescaled variables $X=x/R$, $Y=y/R$. 
This leads to the exact formula \cite{Allegra_2016}
\begin{equation}\label{eq:exactdw}
 \rho(X,Y)=\int_{C_k}\frac{dk}{2\pi} \int_{C_q} \frac{dq}{2\pi}
 \frac{e^{R\left(\varphi(k)-\varphi(q)\right)}}{1-e^{-\ci (k-q)}},
\end{equation}
where 
\begin{equation}
 \varphi(k)=-\ci k X+Y\varepsilon(k)-\ci \tilde{\varepsilon}(k).
\end{equation}
Note that we lightened the notations and removed the dependency on $x,y$ in $\varphi(k)$ compared to the previous section.
The contours are the segments $C_k=[-\pi,\pi]$, $C_q=[-\pi+\ci \eta,\pi+\ci \eta]$ where $\eta>0$. To keep track of which contour corresponds to which variable, we have found it very convenient to put a (otherwise meaningless) subscript indicating this.

Our next aim is to study the asymptotics for $R\to \infty$ with $X\in \mathbb{R}$ and $Y\in (-1,1)$ fixed, which can be done using the steepest descent method (see e.g. \cite{BOO}). We are free to deform both integration contours, as long as their image through $k\mapsto e^{ik}$ defines a smooth contour in the complex plane. This means $k$ lives on an infinite cylinder with circumference $2\pi$. In particular, for each contour the end points with real parts $\pm \pi$ must have the same imaginary part and a horizontal tangent. 
To keep the same value for the integral (\ref{eq:exactdw}), no singularity of the integrand must be encountered in the deformation process: this implies that $C_{q}$ must always be on top of $C_k$ in the complex plane (i.e. given $q\in C_q$ and $k\in C_k$ such that $\re q=\re k$ we have $\im q>\im k$), otherwise we hit the pole at $k=q$.
 
The first step is look for saddle points, which are solutions of 
\begin{equation}\label{eq:spe}
  X+\ci Y\frac{d\varepsilon(k)}{dk}+ \frac{d\tilde{\varepsilon}(k)}{dk}=0.
\end{equation}
Provided one finds the correct paths of steepest descent going through the saddle points, it will then be possible to estimate the integrals asymptotically. Equation (\ref{eq:spe}) is a quartic equation in $\omega=e^{\ci k}$, so can be solved explicitly. To prepare for the general analysis, we find it necessary to study the particular --well known-- case $\alpha=0$ first. This is done in the next subsection.
\subsection{Nearest-neighbor case}
\label{sec:nn}
\begin{figure}[htbp]
\begin{tikzpicture}
 \node at (0,0) {\includegraphics[width=0.24\textwidth]{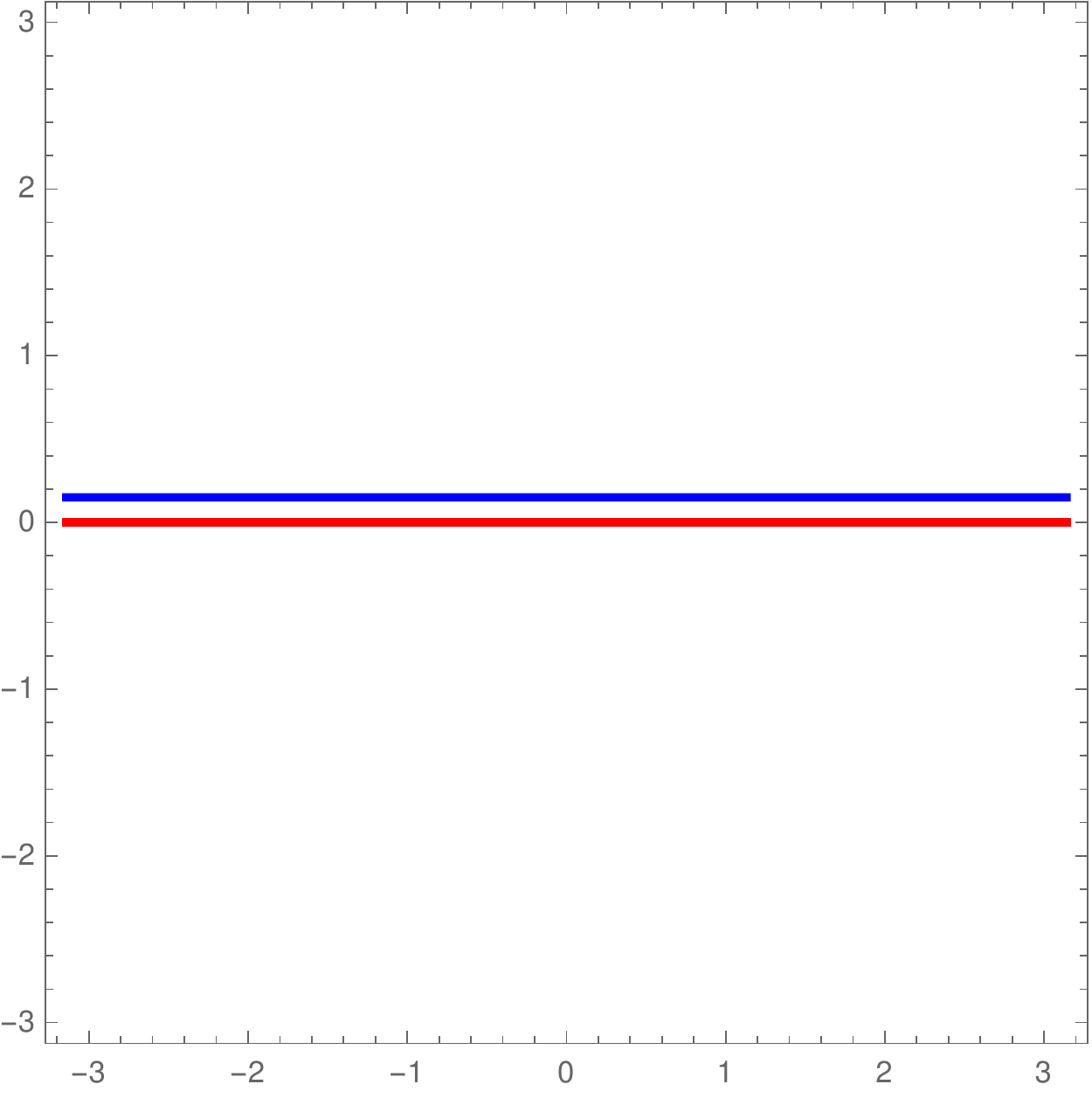}};
 \node at (4.6,0) {\includegraphics[width=0.24\textwidth]{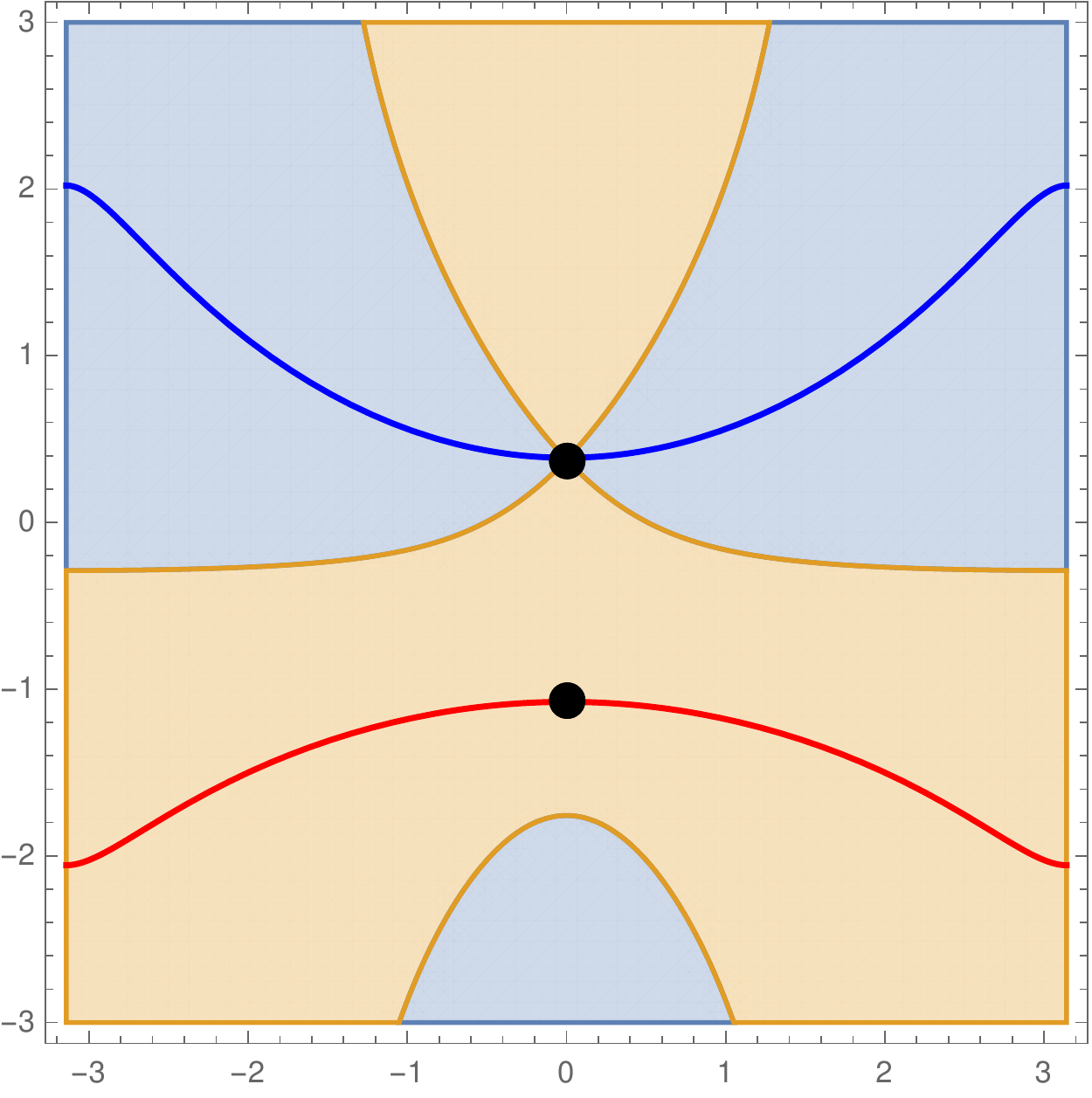}};
 \node at (13.8,0) {\includegraphics[width=0.24\textwidth]{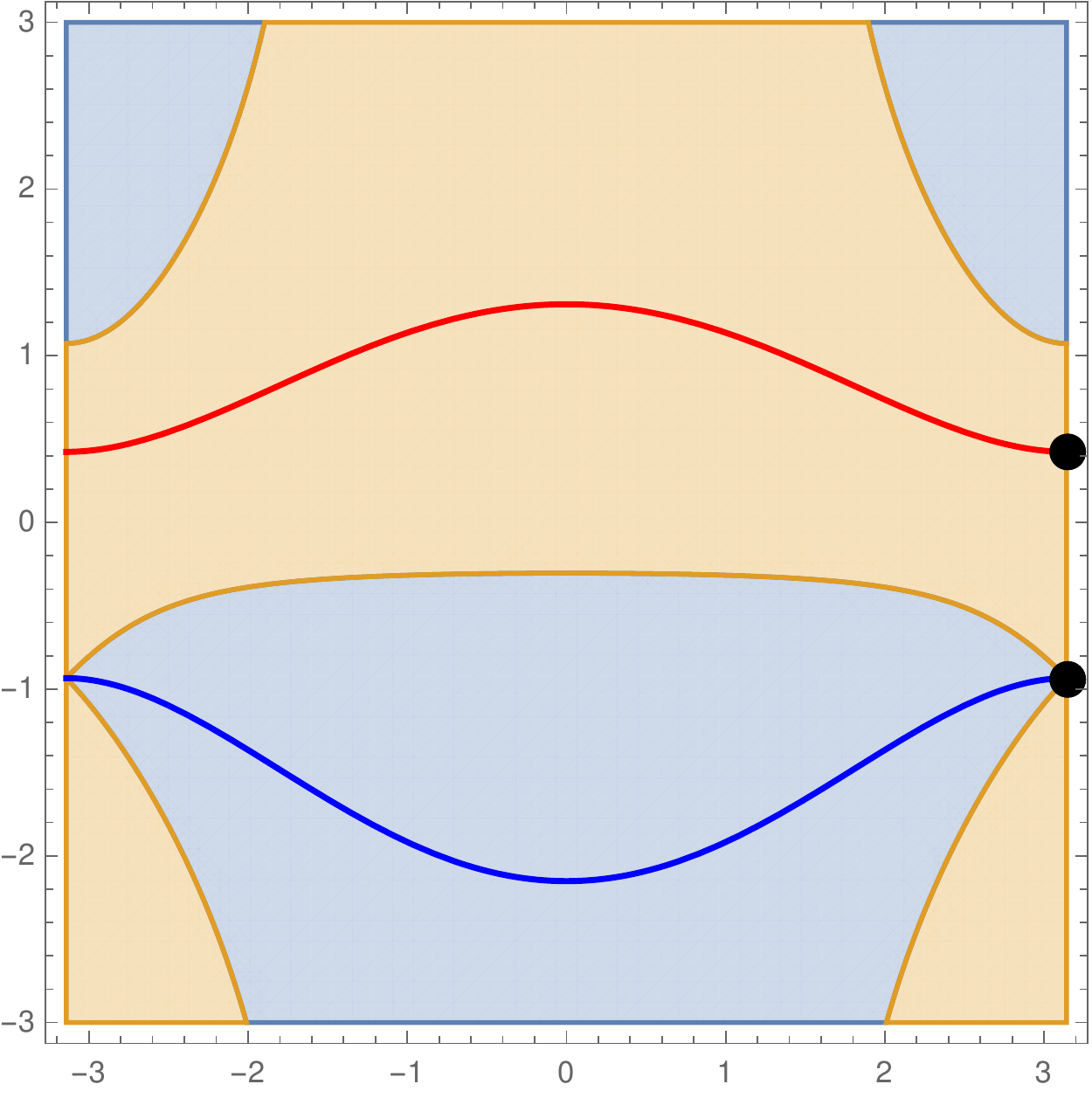}};
 \node at (9.2,0) {\includegraphics[width=0.24\textwidth]{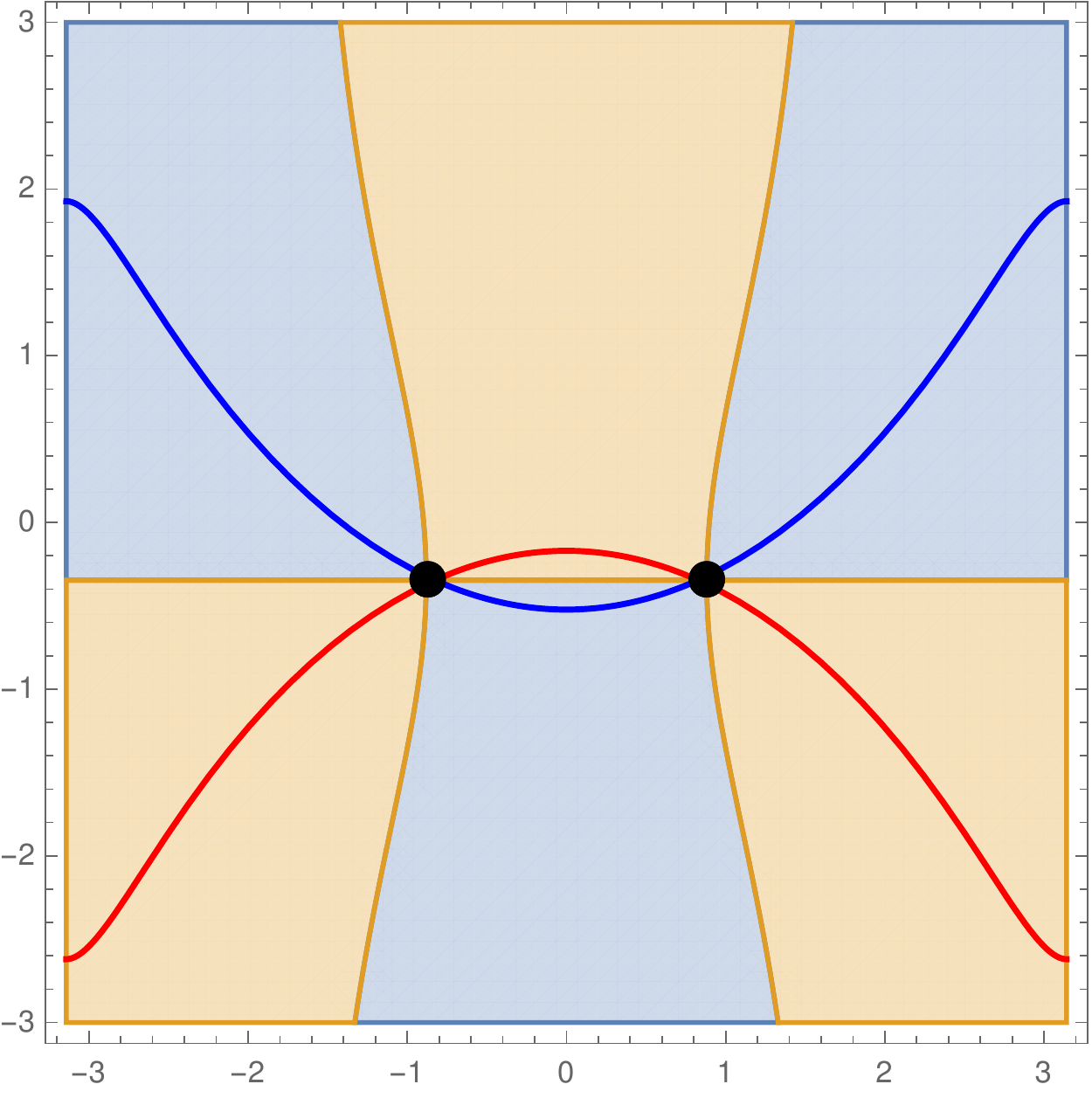}};
 \draw (4.6,2.5) node {(a)\quad $\rho(X,Y)=0$};
 \draw (13.8,2.5) node {(c)\quad $\rho(X,Y)=1$};
 \draw (9.2,2.5) node {(b)\quad $0<\rho(X,Y)<1$};
   \draw[red] (0,-0.2) node {$C_k$};
   \draw[blue] (0,0.5) node {$C_q$};
 \end{tikzpicture}
\caption{Illustration of the saddle point method in the three different regimes. Left: Initial contours $C_k$ (red) and $C_q$ (blue). (a): regime I, with $\Delta>0$ and $X>0$. Black dots are the two saddle points. The blue shaded regions has $\re\varphi(k)>\re\varphi(z_+)$, while the reverse holds in the orange region. The deformations of the contours shown ensure that density decays exponentially fast.  (b): regime II, $\Delta<0$. In that case the two saddle points are image of each others through the $y=0$ axis. To get those contours, it is necessary for $C_k$ and $C_q$ to partially cross on the segment $[ z_-, z_+]$, yielding formula (\ref{eq:densityinside}). (c) regime III, with $\Delta>0$ and $X<0$. The deformation shown also ensures decay to zero, however it is necessary for the two contours to cross to obtain the shown contours. The extra contribution from the residue gives $\braket{n_x(y)}_R\sim 1$.}
\label{fig:contourdeformations}      
\end{figure}
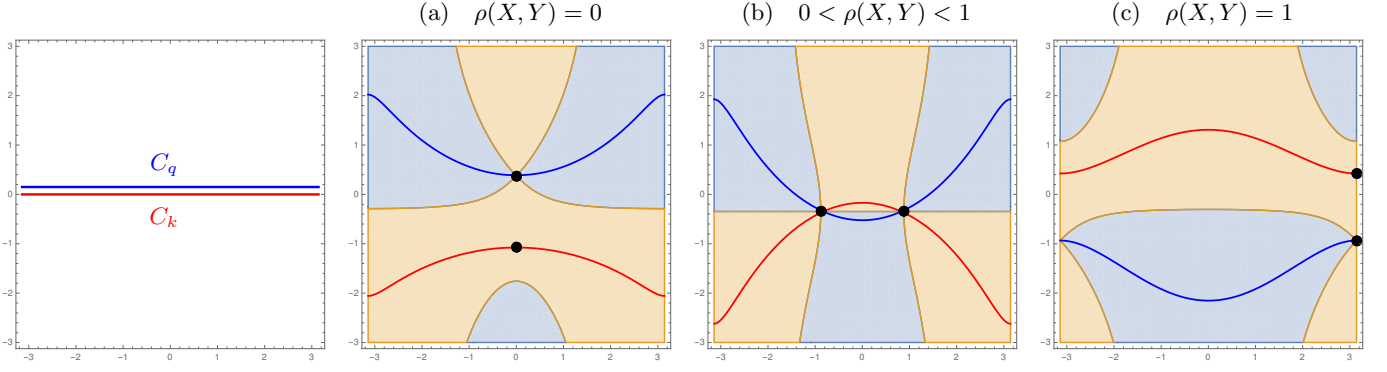

Setting $\omega=e^{\ci k}\in \mathbb{C}\backslash\{0\}$, the saddle point equation reduces to
\begin{equation}\label{eq:omega}
 (1-Y)\omega^2-2X \omega+(1+Y)=0,
\end{equation}
which is a second degree equation in $\omega$, with discriminant
\begin{equation}
 \Delta=4\left(X^2+Y^2-1\right).
\end{equation}
The analysis depends on the sign of the discriminant, and for $\Delta>0$ also on the sign of $X$. This defines three regimes, which we study separately. To help visualize the saddle points and various contour deformation that will follow, we show them in figure \ref{fig:contourdeformations}.
\paragraph{Regime I: $\Delta>0$ and $X>0$.} This is the simplest one. The solutions to (\ref{eq:omega}) are real, which means, since $\omega=e^{\ci k}$, that the solutions to the saddle point equation (\ref{eq:spe}) are pure imaginary:
\begin{equation}
 z_{\pm}=\ci\, \textrm{arctanh}\left(\frac{-Y\pm  X\sqrt{X^2+Y^2-1}}{X^2+Y^2}\right).
\end{equation}
Now, we deform $C_k$ to pass through $z_-$ and $C_q$ to pass through $z_+$. No crossing occurs since $z_+$ has largest imaginary part. Finding the path of steepest descent, the integrals are dominated by the neighborhood of the saddle points. The fact that $\varphi(z_-)-\varphi(z_+)$ is real and negative ensures exponential decay to zero. It is possible to be more precise, Taylor expanding $\varphi$ about both saddle points. We obtain
\begin{align}
 \rho(X,Y)&\sim \frac{1}{2\pi R\sqrt{\phi''(z_-)\phi''(z_+)}}\frac{e^{R(\varphi(z_-)-\varphi(z_+))}}{2\ci \sin\left(\frac{z_--z_+}{2}\right)}\\
 &\sim\frac{\sqrt{1-Y^2}}{4\pi R(X^2+Y^2-1)}\exp\left(-2R\left[X\textrm{arctanh} \left(\frac{\sqrt{X^2+Y^2-1}}{X}\right)-\sqrt{X^2+Y^2-1}\right]\right),
\end{align}
which decays exponentially since $\textrm{arctanh}\, u>u$ for $u>0$. 

It is not strictly necessary to find the path of steepest descent, if one just want to show exponential decay to zero without computing the decay rate or the prefactor. Indeed by trivial bounds, we have
\begin{equation}
\left| \int \frac{dk dq}{(2\pi)^2} \frac{e^{R(\varphi(k))-\varphi(q))}}{1-e^{-\ci (k-q)}}\right|
\leq \int \frac{dk dq}{(2\pi)^2} \frac{e^{R(\re \varphi(k)-\re\varphi(q))}}{2\left|\sin\frac{k-q}{2}\right|}.
\end{equation}
To show that the rhs decays to zero, it is sufficient to look at the level lines of $\varphi(z_+)$ as illustrated in figure \ref{fig:contourdeformations}(a). We show in light blue the region for which $\re \varphi(k)>\re\varphi(z_+)$, and in light orange the region $\re \varphi(k)<\re\varphi(z_+)$. Then, deforming $C_q$ to \emph{any} contour in the blue region, and $C_k$ to \emph{any} contour in the orange region ensures that $\re (\varphi(k)-\varphi(q))<\textrm{cst}<0$. Hence we are integrating a function that is exponentially small everywhere, so the density itself is. This is possible here without encountering any singularity, so the result $\rho(X,Y)\to 0$ follows. We handle the next two regimes using this last simpler method.
\paragraph{Regime II: $\Delta<0$.} This regime is slightly subtler, since the stationary points become full complex numbers \cite{Allegra_2016}:
\begin{equation}
 z_{\pm} =\pm \arccos \frac{X}{\sqrt{1-Y^2}}-\ci \textrm{arctanh} Y.
\end{equation}
The contours ensuring decay to zero are shown in figure \ref{fig:contourdeformations}(c). It is possible to deform those according to the previous prescription, but not without having $C_k$ and $C_q$ cross partially during the process. Choosing this crossing to occur along the segment $[z_-,z_+]$, we get an extra residue contribution
\begin{align}
  \rho(X,Y)&=\int_{-\pi}^{\pi}\frac{dk}{2\pi} \int_{-\pi+\ci \eta}^{\pi+\ci \eta} \frac{dq}{2\pi}\frac{e^{R\left(\varphi(k)-\varphi(q)\right)}}{1-e^{-\ci (k-q)}}\\
  &=\int_{C'_k}  \frac{dk}{2\pi}\int_{C'_q}\frac{dq}{2\pi}\frac{e^{R\left(\varphi(k)-\varphi(q)\right)}}{1-e^{-\ci (k-q)}}+\int_{k_-}^{k_+} \frac{dk}{2\pi},
\end{align}
where $C'_k, C'_q$ are the contours shown in figure \ref{fig:contourdeformations}(b). The first term on the rhs is a subleading power correction since in the neighborhood of $z_{\pm}$ $C_k$ and $C_q$ are very close. This is very much expected from a critical system, see e.g \cite{Allegra_2016}. The second term therefore dominates, and we obtain
\begin{equation}\label{eq:densityinside}
 \rho(X,Y)\sim \frac{z_+-z_-}{2\pi}=\frac{1}{\pi}\arccos \frac{X}{\sqrt{1-Y^2}}.
\end{equation}
The density in the fluctuating region is, in the end, simply the length of the segment $[z_-,z_+]$, divided by $2\pi$. Said differently, $\rho(X,Y)\sim \frac{\re z_+}{\pi}$, so the real part of $z_+$ plays the role of Fermi momentum in standard ground state quantum mechanics. The imaginary part also has an interpretation \cite{Abanov_hydro,Stephan_ebc}, related to the current. 

\paragraph{Regime III: $\Delta>0$ and $X<0$.}
The last regime simply follows from the symmetry $\rho(-X,Y)=1-\rho(X,Y)$, which can easily be shown to be exact at the lattice level. It is however instructive to recover this from the saddle point method. Compared to the analysis in regime I, the $z_\pm$ are shifted by $\pi$. It is possible to deform $C_{k}$, $C_q$ in such a way that $C_q$ passes through the stationary point with largest imaginary value, but not without having $C_k$ and $C_q$ fully crossing each other during the process, as illustrated in figure \ref{fig:contourdeformations}(c). In that case one gets an extra contribution stemming from the residue at $k=q$, 
\begin{align}
  \rho(X,Y)= \int_{-\pi+\ci \eta}^{\pi+\ci \eta}  \frac{dk}{2\pi}\int_{-\pi}^{\pi}\frac{dq}{2\pi}\frac{e^{R\left(\varphi(k)-\varphi(q)\right)}}{1-e^{-\ci (k-q)}}+ \int_{-\pi}^{\pi} \frac{dk}{2\pi},
\end{align}
similar to regime II. 
Now the first term on the rhs can be shown to decay exponentially to zero by the argument explained for regime I, while the second equals one. Hence the density is one up to exponentially small corrections in region III.

Note finally that the limit case $\Delta=0$ which we excluded from the analysis corresponds to the arctic curve, which is the circle $X^2+Y^2=1$ here. Close to the circle, density approaches $0$ or $1$ as square root, due to the fact that the two saddle points coalesce.

\subsection{The general case}
\label{sec:general}
Let us now assume $\alpha>0$. In terms of $\omega=e^{ik}$ the saddle point equation reads
\begin{equation}
 \label{eq:spe_full}
 2\alpha (1-Y)\omega^4+(1-Y)\omega^3-2X\omega^2+(1+Y)\omega+2\alpha(1+Y)=0,
\end{equation}
a quartic equation with four roots, including possible multiplicities. There are three classes of solutions to the above equation. Either (i) all solutions are real, or (ii) two solutions are real, and two are complex conjugated, or (iii) all four solutions are complex, in which case they come in two pairs of complex-conjugated roots. There is still a discriminant $\Delta$ associated to such polynomial equations. $\Delta<0$ corresponds to (ii), while $\Delta>0$ corresponds to (i) or (iii). It may be computed explicitly, as shown below
\begin{equation}
 \Delta=-32 \alpha  X^3 (8 \alpha  X+1)+\left(Y^2-1\right) \left[-108 \alpha ^2+\left(1-64 \alpha ^2 \left(32 \alpha ^2+5\right)\right) X^2-36 \left(32 \alpha ^3+\alpha \right) X\right]-\left(16 \alpha ^2-1\right)^3 \left(Y^2-1\right)^2.
\end{equation}
We have removed a trivial multiplicative factor $4(Y^2-1)$, which cannot vanish. The above expression is not particularly illuminating, except in a few selected cases. 

\begin{figure}[htbp]
\begin{tikzpicture}
 \node at (0,5.5) {\includegraphics[width=0.25\textwidth]{./figures/XX_initcontour.pdf}};
 \node at (6.5,5.5) {\includegraphics[width=0.25\textwidth]{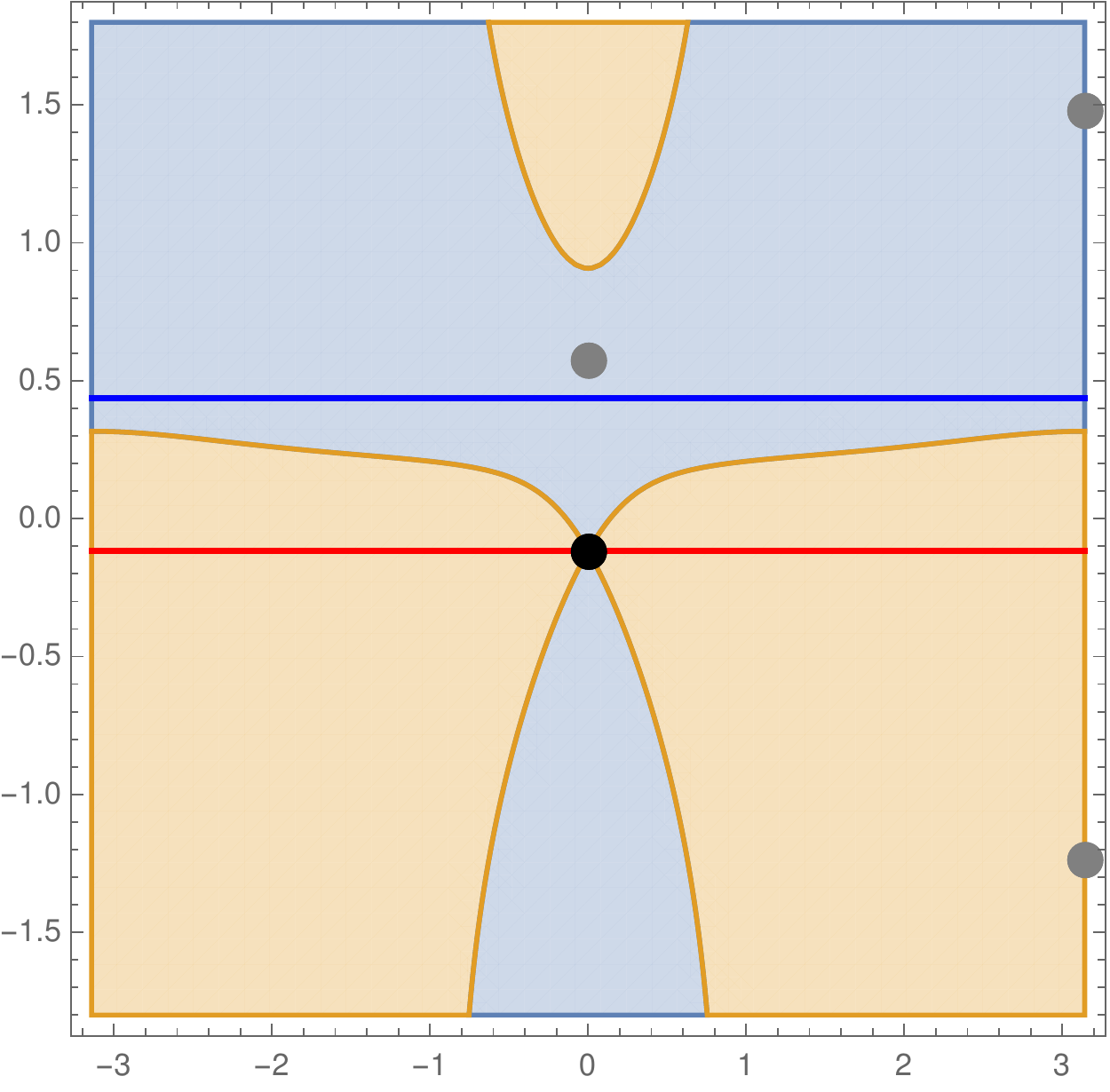}};
 \node at (13,5.5) {\includegraphics[width=0.25\textwidth]{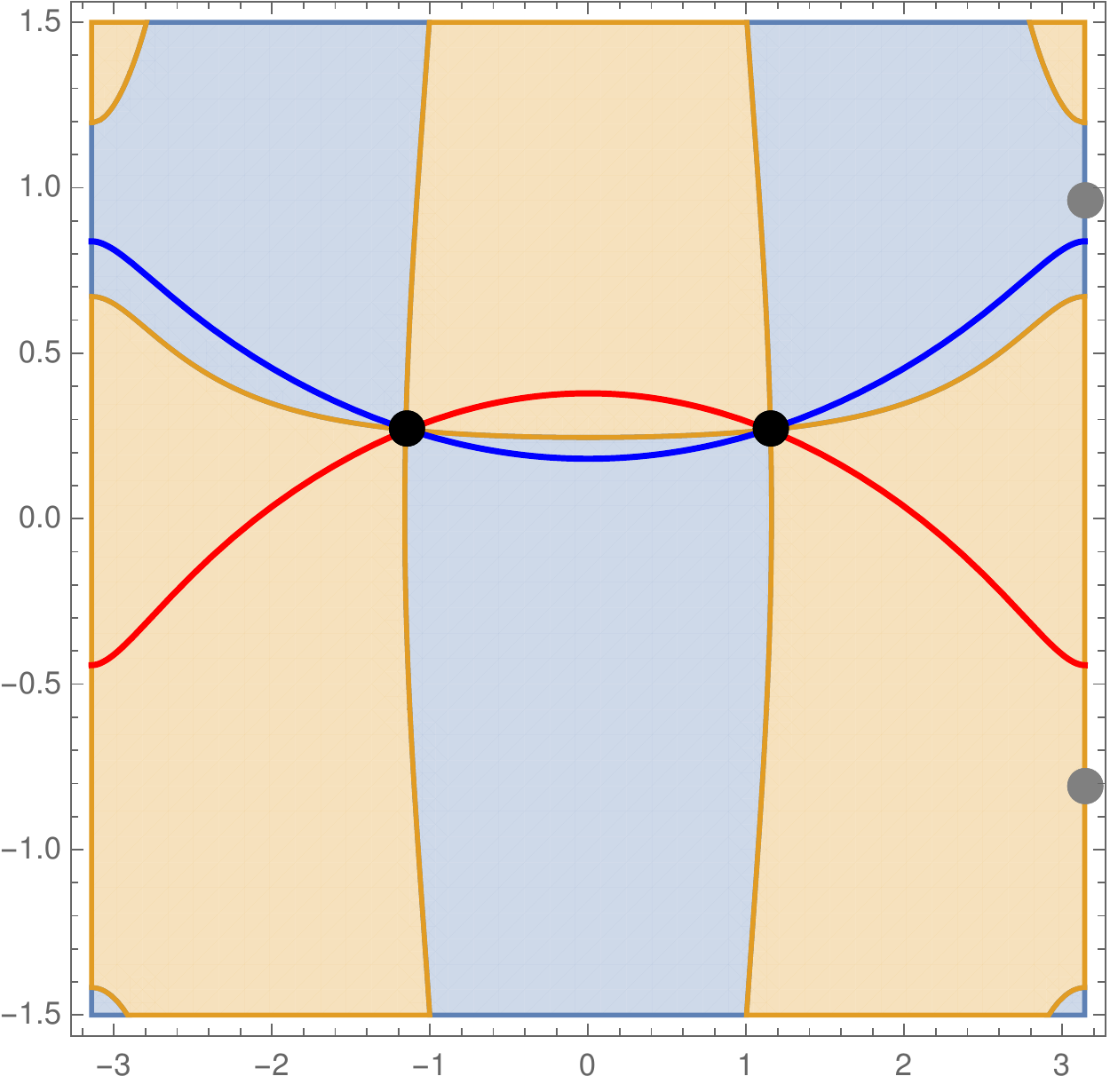}};
  \node at (0,0) {\includegraphics[width=0.25\textwidth]{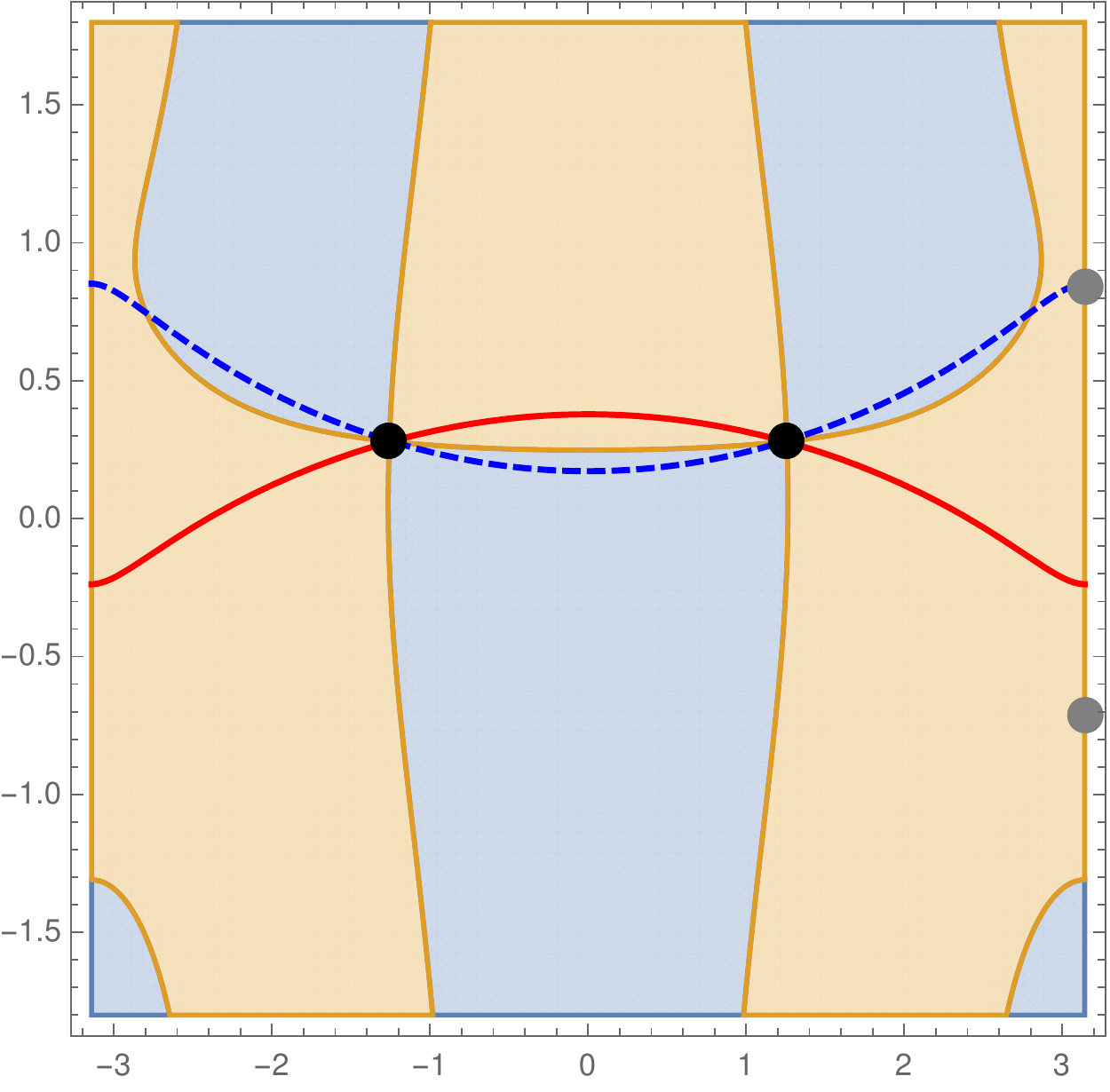}};
 \node at (6.5,0) {\includegraphics[width=0.25\textwidth]{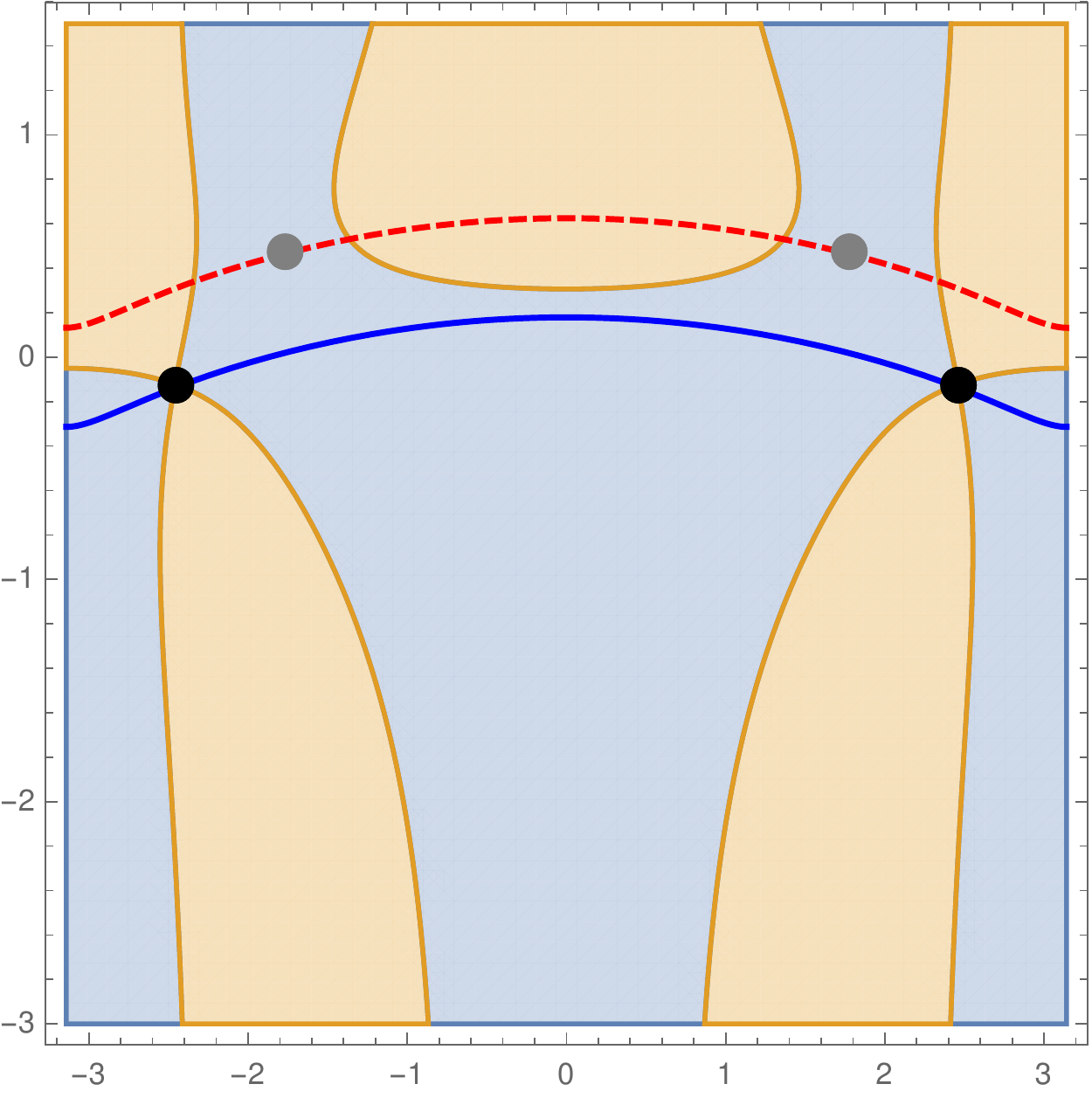}};
 \node at (13,0) {\includegraphics[width=0.25\textwidth]{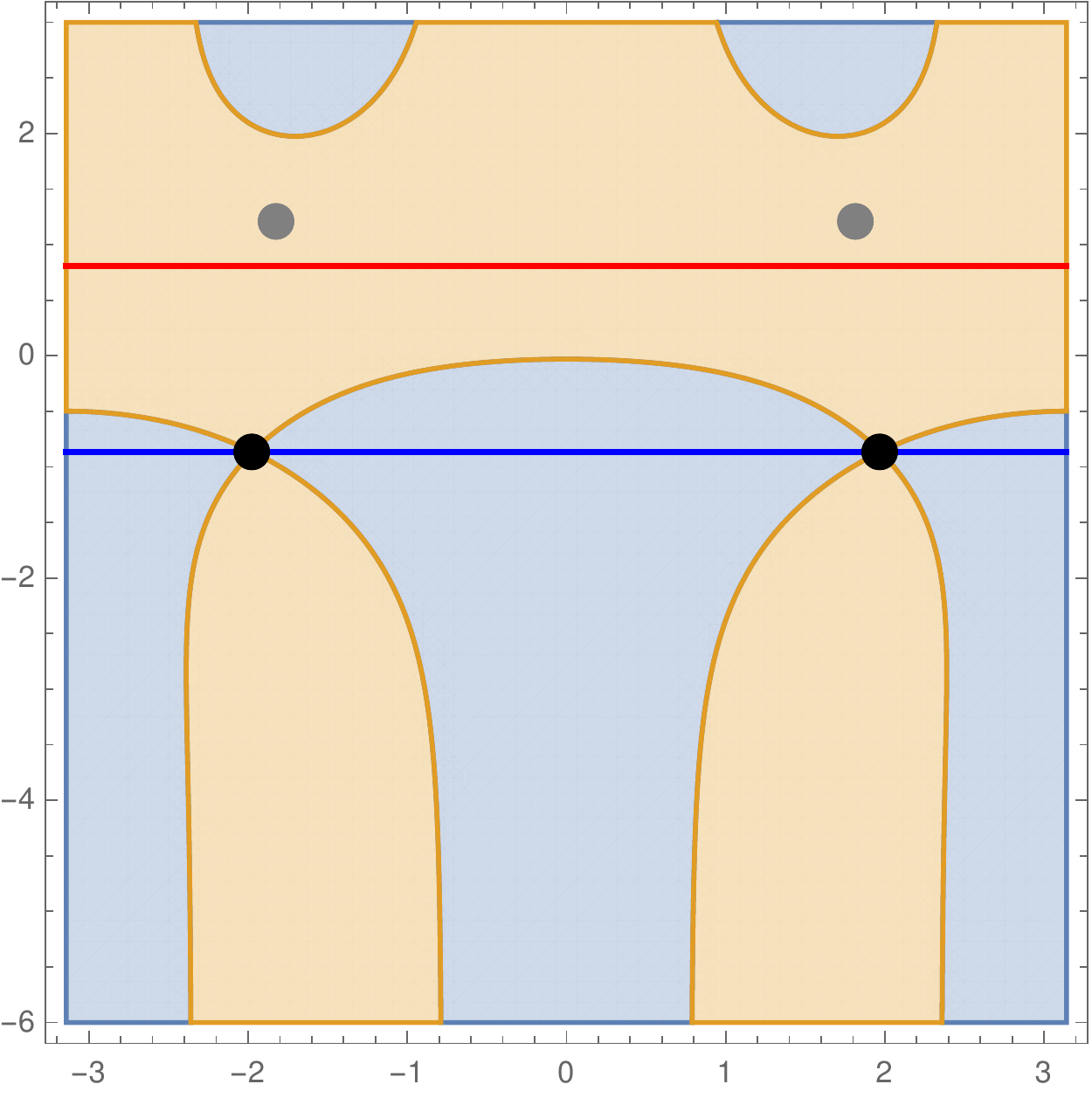}};
 \draw (6.5,8) node {(a)\quad $\rho(X,Y)=0$};
  \draw (13,8) node {(b)\quad $0<\rho(X,Y)<1$};
 \draw (0,2.5) node {(c)\quad $|\rho(X,Y)|\to \infty$};
  \draw (6.5,2.5) node {(d)\quad $|\rho(X,Y)|\to \infty$};
  \draw (13,2.5) node {(e)\quad $\rho(X,Y)= 1$};
  \draw[red] (0,5.35) node {$C_k$};
   \draw[blue] (0,5.95) node {$C_q$};
\end{tikzpicture}
 \caption{Illustration of the various contour deformations, on the example $Y=-1/3$ and $\alpha=1/4$. Top left: initial contours. (a) $X=1.6$. Regime I, frozen region with density zero. (b) $X=0.06$. Regime II, fluctuating region with density (\ref{eq:normal}). (c) $X=-0.1$. Regime III, crazy region. (d) $X=-0.7$. Regime IV, crazy region. (e) $X=-2$. Regime V, frozen region with density one.}
 \label{fig:alphacontours}
\end{figure}

The whole behavior of roots turns out to also depend on $\alpha$. We present an analysis for $\alpha>1/8$, which is the least favorable from a positivity perspective. This will be sufficient to understand the mechanism responsible for the appearance of crazy regions, sidestepping a general and laborious treatment. We identified five possible regimes, which are all illustrated in figure \ref{fig:alphacontours}. As before the regimes go from the far east (I) to the far west (V). For the whole analysis, we label the stationary points as $z_1$, $z_2$, $z_3$, $z_4$ and order them from smallest to biggest real part, and if two real parts coincide from smallest to biggest imaginary part. To lift possible ambiguities, we take $\re z_i \in (-\pi,\pi]$.
\paragraph{Regime I:} In this regime, all the $e^{\ci z_i}$ are real, two positive, two negative. In terms of the $z_i$, this means two solutions have real part $0$, while two have real part $\pi$. It is sufficient to study the sign of $\re[\varphi(k)-\varphi(z_1)]$ to evaluate the asymptotic value of the integral: denoting as before in blue the region $\re\varphi(k)>\re\varphi(z_1)$, and orange the region $\re\varphi(k)<\re\varphi(z_1)$, it is possible to deform $C_q$ to the blue region, while deforming $C_k$ to the orange region without hitting any singularity. Hence 
\begin{equation}
 \rho(X,Y)\to 0
\end{equation}
exponentially fast. This is illustrated in figure \ref{fig:alphacontours}(a). We are in the east frozen region.
\paragraph{Regime II:} In this regime $z_2=-z_1^*$, while $z_3$, $z_4$ have real part $\pi$, and the level lines of $\re \varphi(z_1)=\re \varphi(z_2)$ have the feature that it is possible to draw a contour going only in the blue region, and for the orange region as well. See figure \ref{fig:alphacontours}(b) for an example. When this holds the other solutions $z_3$ and $z_4$ play no role. The right deformation can be achieved at the price of having $C_k$ and $C_q$ partially cross each other, resulting in an extra residue contribution. We obtain
\begin{equation}\label{eq:normal}
 \rho(X,Y)\sim \frac{1}{2\pi} \left(z_2-z_1\right)
\end{equation}
with power law subleading corrections. This is very similar to regime II for $\alpha=0$ in the previous subsection. We are in a fluctuating region, with a well-behaved density. $\re z_2$ plays the role of Fermi momentum.
\paragraph{Regime III:} In this regime $z_2=-z_1^*$, while $z_3$, $z_4$ have real part $\pi$ as before. However, it is not possible to draw a contour solely in the blue region anymore. 
One can show that we cannot perform the deformation as desired even considering the other saddle points as the benchmark to shade the regions in the picture. We can use the same approach as in regime II, with a residue contribution identical to (\ref{eq:normal}). However, the other contribution has no reason to be exponentially small anymore, since one cannot avoid $\re[\varphi(k)-\varphi(q)]>0$ somewhere along the contours. This is illustrated in figure \ref{fig:alphacontours}(c). Then, the dominant contribution is obtained by having $C_q$ go through $z_4$ and keeping $C_k$ going through $z_1,z_2$. We obtain 
\begin{equation}
 \rho(X,Y)\sim \frac{1}{2\pi R \sqrt{\varphi''(z_4)}} \left[\frac{e^{R(\varphi(z_1)-\varphi(z_4))}}{2\ci \sqrt{\varphi''(z_1)}\sin \left(\frac{z_1-z_4}{2}\right)}
+\frac{e^{R(\varphi(z_2)-\varphi(z_4))}}{2\ci \sqrt{\varphi''(z_2)}\sin \left(\frac{z_2-z_4}{2}\right)}\right],
\end{equation}
which can be shown to be real.
Since $\re[\varphi(z_{1,2})-\varphi(z_4)]>0$, we get a sum of two terms which blow up exponentially in modulus, meaning the density diverges almost everywhere in this regime. This is an instance of a crazy region.
\paragraph{Regime IV: } In this regime all roots are complex, with $z_2=-z_1^*$, $z_4=-z_3^*$. We run into a similar problem as in Regime III, which means the density also diverges almost everywhere. See figure \ref{fig:alphacontours}(d). We deform $C_k$ to go through $z_2$, $z_3$, while $C_q$ goes through $z_1$, $z_4$. This is also a crazy region.
\paragraph{Regime V:} In this regime all roots are complex, but it is easy to find a deformation which decays to zero, at the expense of fully exchanging $C_k$ and $C_q$. The corresponding residue contribution gives $\rho(X,Y)=1$ up to exponentially small corrections, we are in the frozen region on the west.

Let us finish this analysis with general comments regarding the emergence of crazy regions. The reason why those may appear is simply related to the fact that we have more than two (anticonjugated) saddle points.  For $\alpha=0$ those define two valleys which have a simple structure, and both valleys wrap around the cylinder. Preventing this wrapping  property requires the existence of more saddle points, so, when there are more, guaranteeing  $\re(\varphi(k)-\varphi(q))\leq 0$ along both contours might not be possible anymore. We then happen to always be able to find a region in parameter space where the integral is dominated by a region where $\re(\varphi(k)-\varphi(q))>0$, and the imaginary part is not able to fully cancel this exponential blowup.

The analysis is similar for other values of $\alpha$, so let us present the conclusions we reached.  
First, for $\alpha>1/8$, it can be shown that all values of $Y\in (-1,1)\backslash\{0\}$ there are values of $X$ in crazy regions. This is not so for $\alpha<1/8$, where values of $Y$ close to $Y=0$ are well behaved, consistent with figure \ref{fig:densityprofiles} in the introduction. Hence irrespective of $\alpha$, there are always normal and crazy regions. The critical value $\alpha=1/8$ will also play a role in the next subsection.

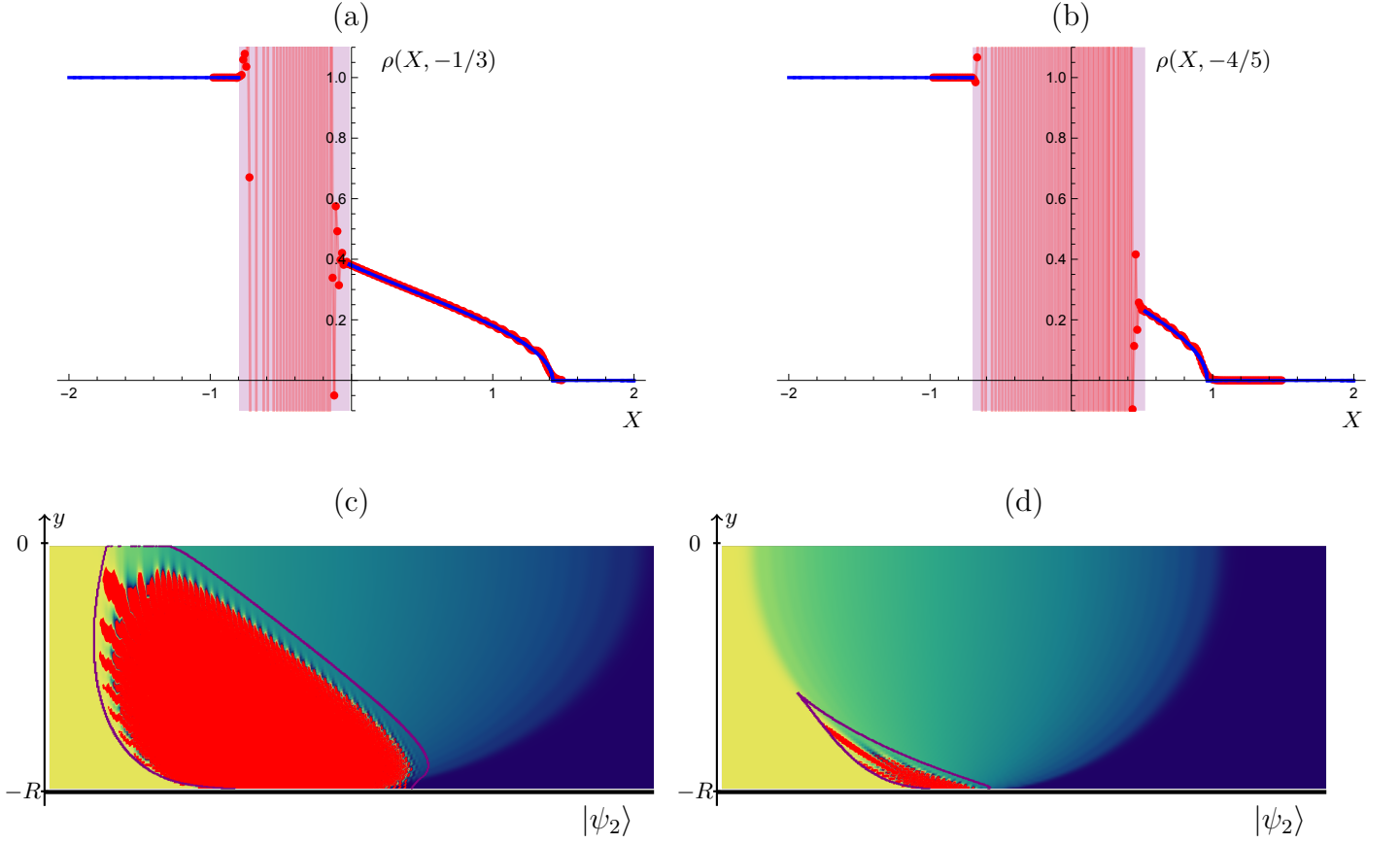
\begin{figure}[htbp]
\begin{tikzpicture}
 \node at (0,0) {\includegraphics[width=0.45\textwidth]{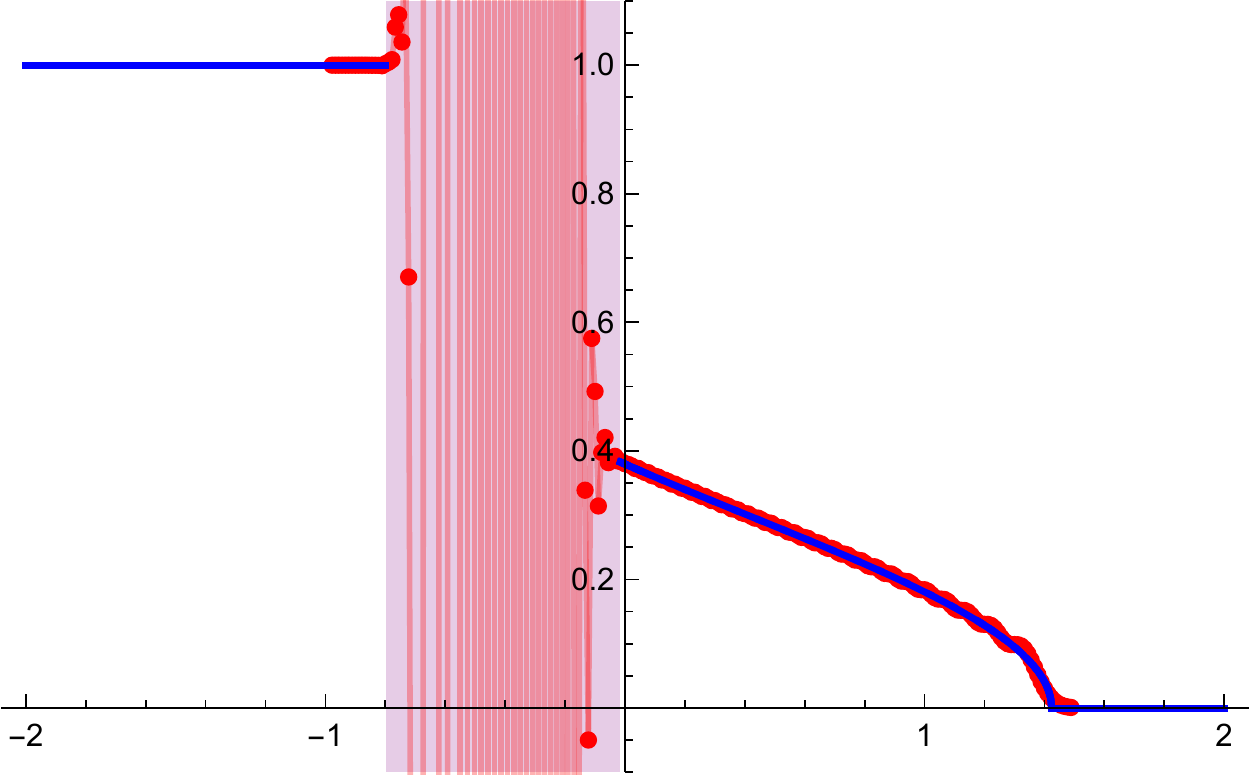}};
 \node at (9.85,0)
 { \includegraphics[width=0.45\textwidth]{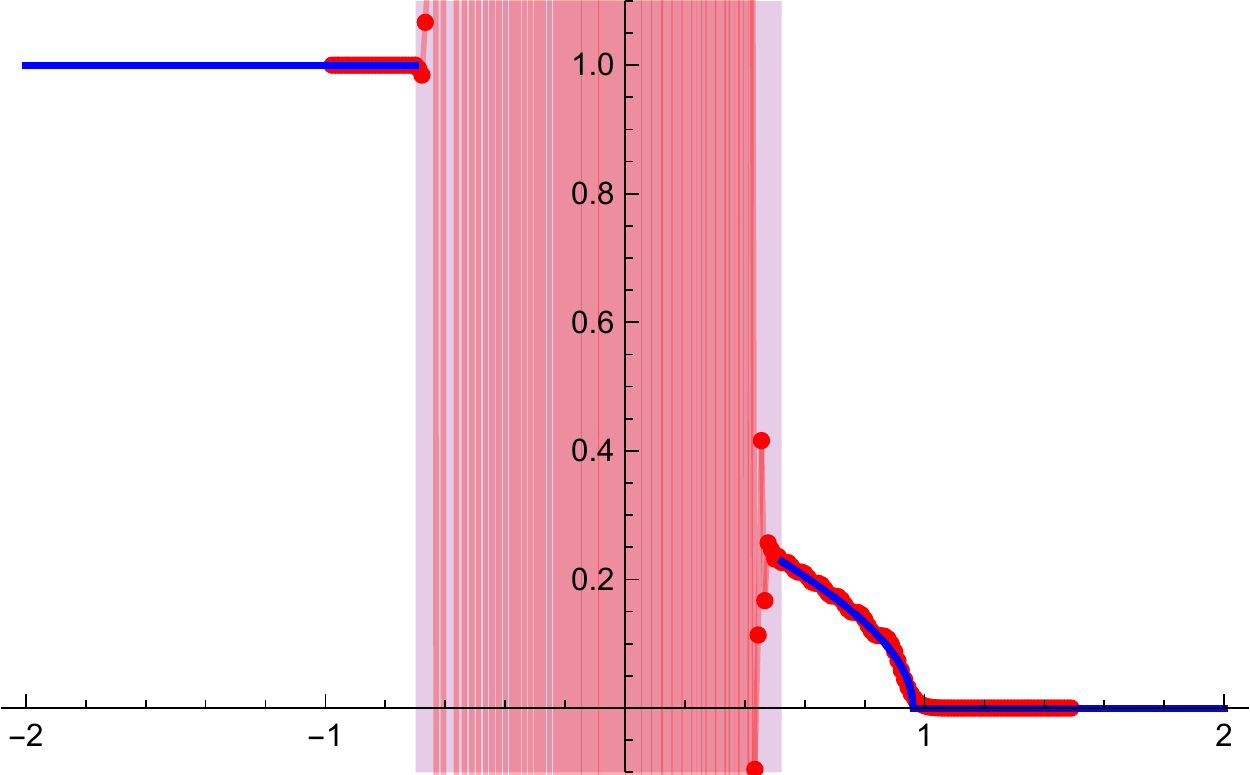}};
 \draw (3.85,-2.6) node {$X$};
 \draw (1.2,2.3) node {$\rho(X,-1/3)$};
  \draw (13.7,-2.6) node {$X$};
 \draw (11.8,2.3) node {$\rho(X,-4/5)$};
 \draw (0,2.9) node {\large{(a)}};
  \draw (9.85,2.9) node {\large{(b)}};
  \begin{scope}[yshift=-6cm]
   \node at (0,0)
 {\includegraphics[width=0.48\textwidth]{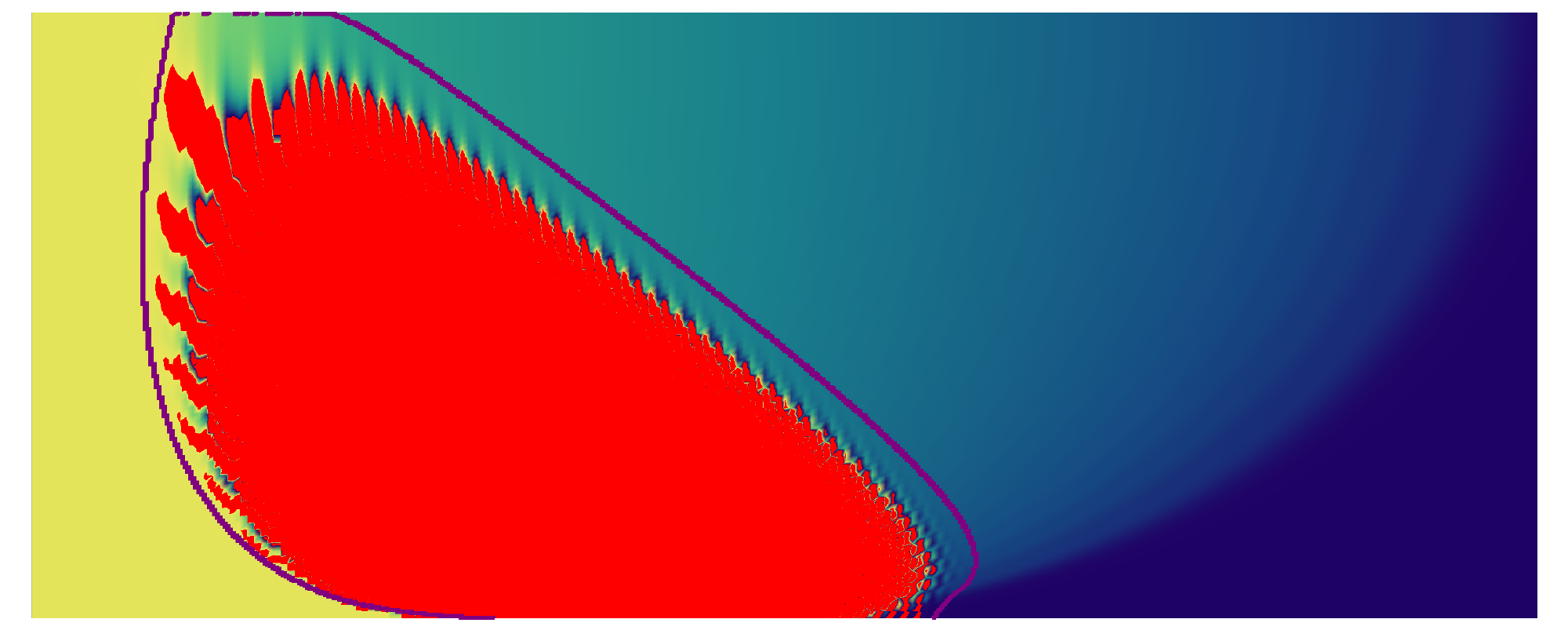}};
 \node at (9.2,0)
 {\includegraphics[width=0.48\textwidth]{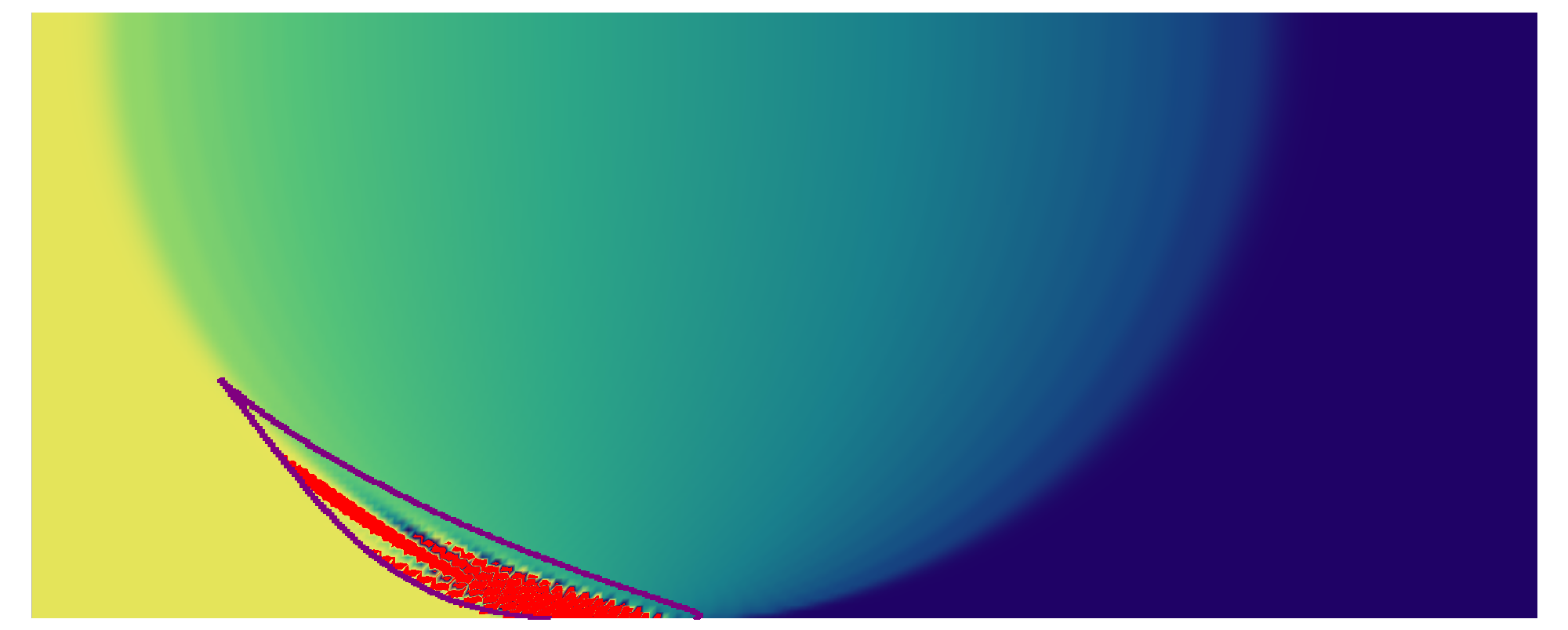}};
  \draw[ultra thick] (-4.2,-1.71) -- (4.14,-1.71);
\draw (3.5,-2.1) node {\large{$\ket{\psi_2}$}};
\draw[thick,->] (-4.2,-1.9) -- (-4.2,2.1);
\draw[thick] (-4.25,-1.7) -- (-4.15,-1.7);
\draw[thick] (-4.25,1.7) -- (-4.15,1.7);
\draw (-4.,2.) node {$y$};
\draw (-4.5,-1.7) node {$-R$};
\draw (-4.5,1.7) node {$0$};
\draw (0,2.25) node {\large{(c)}};
\begin{scope}[xshift=9.2cm]
   \draw[ultra thick] (-4.2,-1.71) -- (4.14,-1.71);
\draw (3.5,-2.1) node {\large{$\ket{\psi_2}$}};
\draw[thick,->] (-4.2,-1.9) -- (-4.2,2.1);
\draw[thick] (-4.25,-1.7) -- (-4.15,-1.7);
\draw[thick] (-4.25,1.7) -- (-4.15,1.7);
\draw (-4.,2.) node {$y$};
\draw (-4.5,-1.7) node {$-R$};
\draw (-4.5,1.7) node {$0$};
\draw (0,2.25) node {\large{(d)}};
\end{scope}
  \end{scope}

\end{tikzpicture}
 \caption{Top: density profiles along slices with constant $Y$ for $\alpha=1/4$ and $R=90$. (a): $Y=-1/3$.  (b): $Y=-4/5$. The numerical result (red dots) is compared to the analytical solution in normal regions, with perfect agreement. The crazy region is shown in shaded violet. Bottom: Bottom-half part of the numerical density profile for $\alpha=1/4$ (c) and $\alpha=1/20$ (d). We use the same color code as in  Fig.~\ref{fig:densityprofiles}, with numerical crazy region in neon red. In addition, the extension of the analytical crazy region is shown in thick violet. The latter two agree well with a small discrepancy which we attribute to finite-size effects.}
 \label{fig:check}
\end{figure}

Several numerical checks of our saddle point analysis are presented in figure \ref{fig:check}, and show excellent agreement. 

It is also worth noting that oscillations we obtain in the CR are very much expected from conservation of the number of particles, which imposes 
\begin{equation}
 \lim_{\Lambda\to \infty} \int_{-\Lambda}^{\Lambda}\left[\rho(X,Y)-\frac{1}{2}\right]dX=0.
\end{equation}
Hence, if the density is very large and positive in some region, it has to be very large in absolute value and negative in some other region, to satisfy the above sum rule.
\subsection{The probabilistic line $Y=0$}
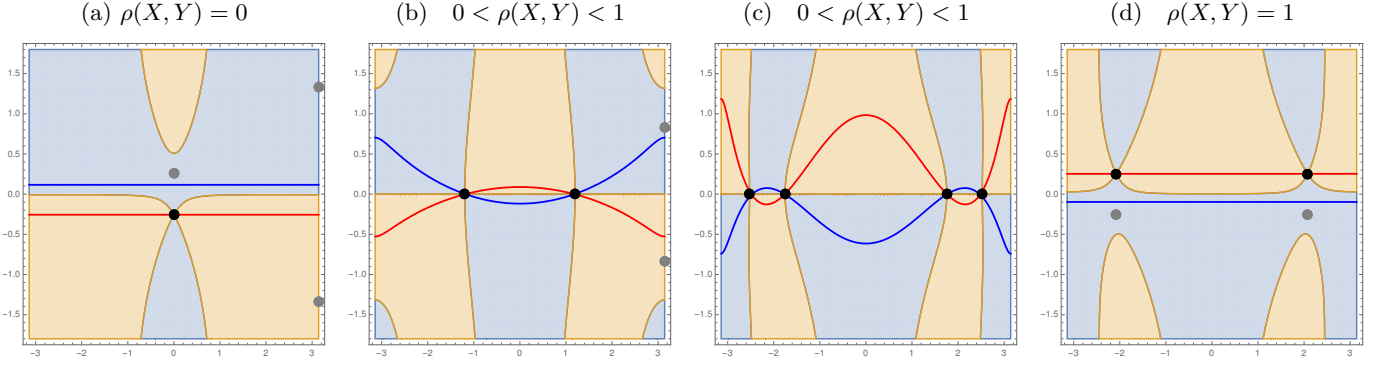
\begin{figure}[htbp]
\begin{tikzpicture}
 \node at (0,0) {\includegraphics[width=0.24\textwidth]{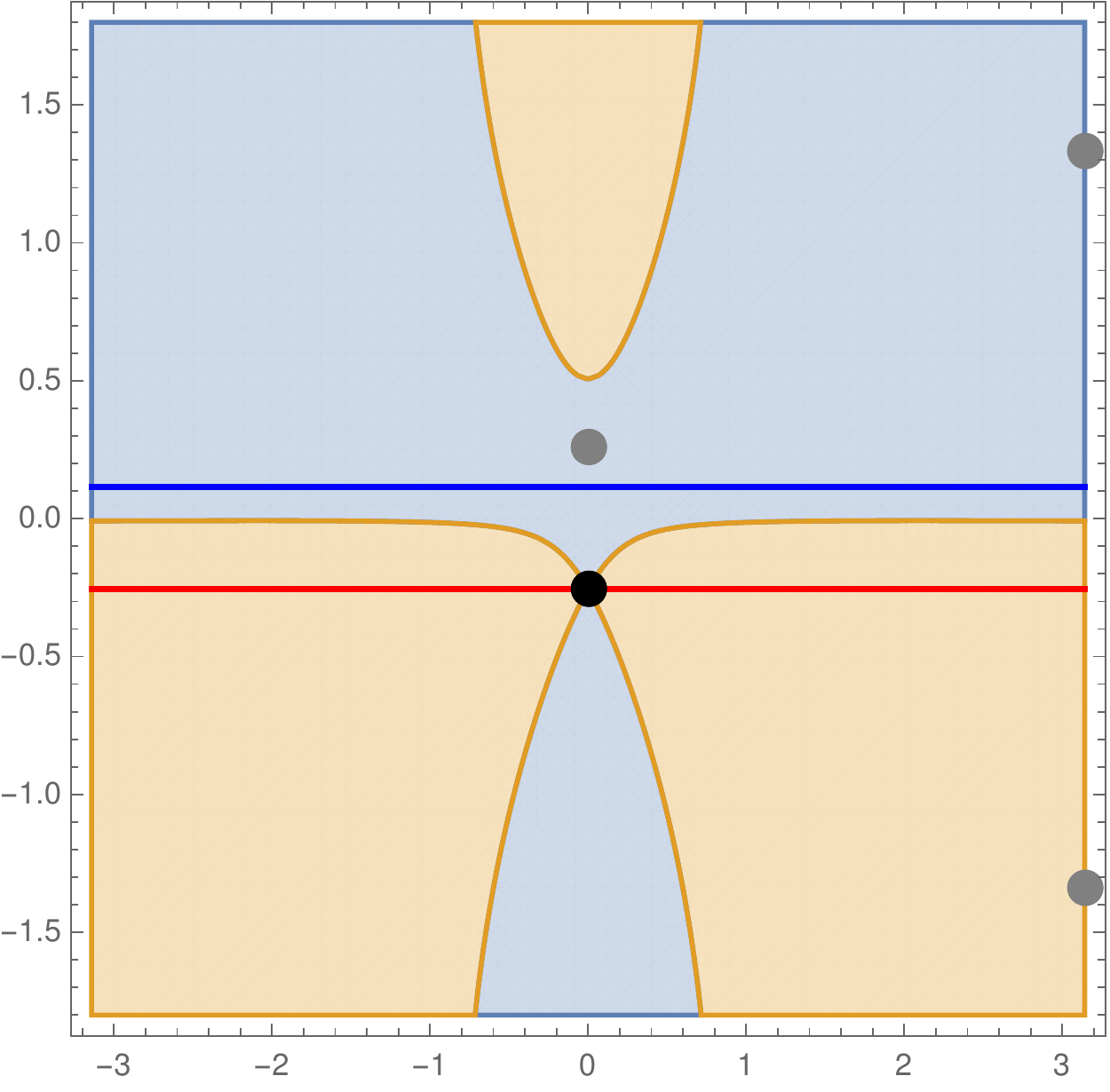}};
 \node at (4.6,0) {\includegraphics[width=0.24\textwidth]{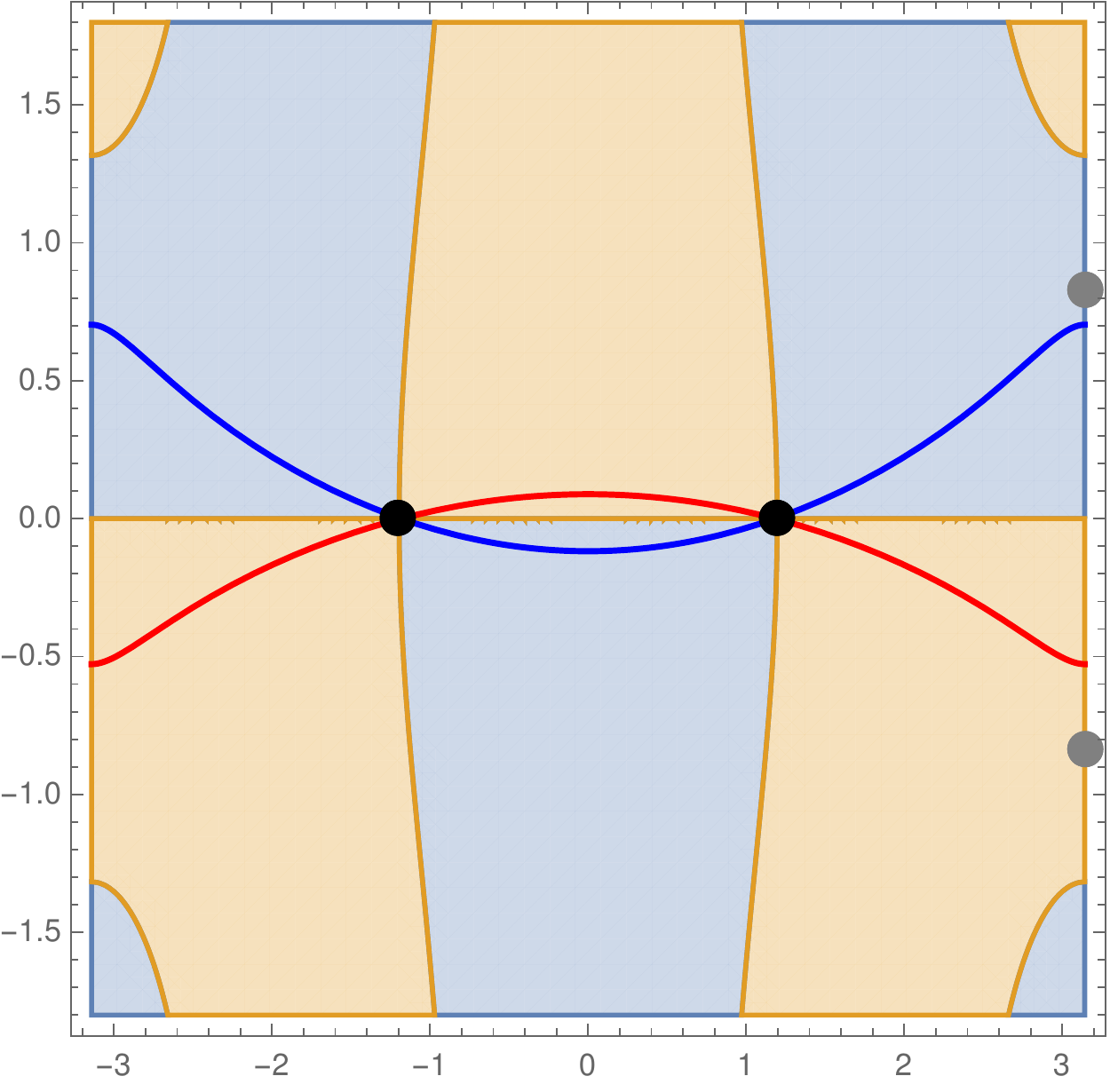}};
 \node at (9.2,0) {\includegraphics[width=0.24\textwidth]{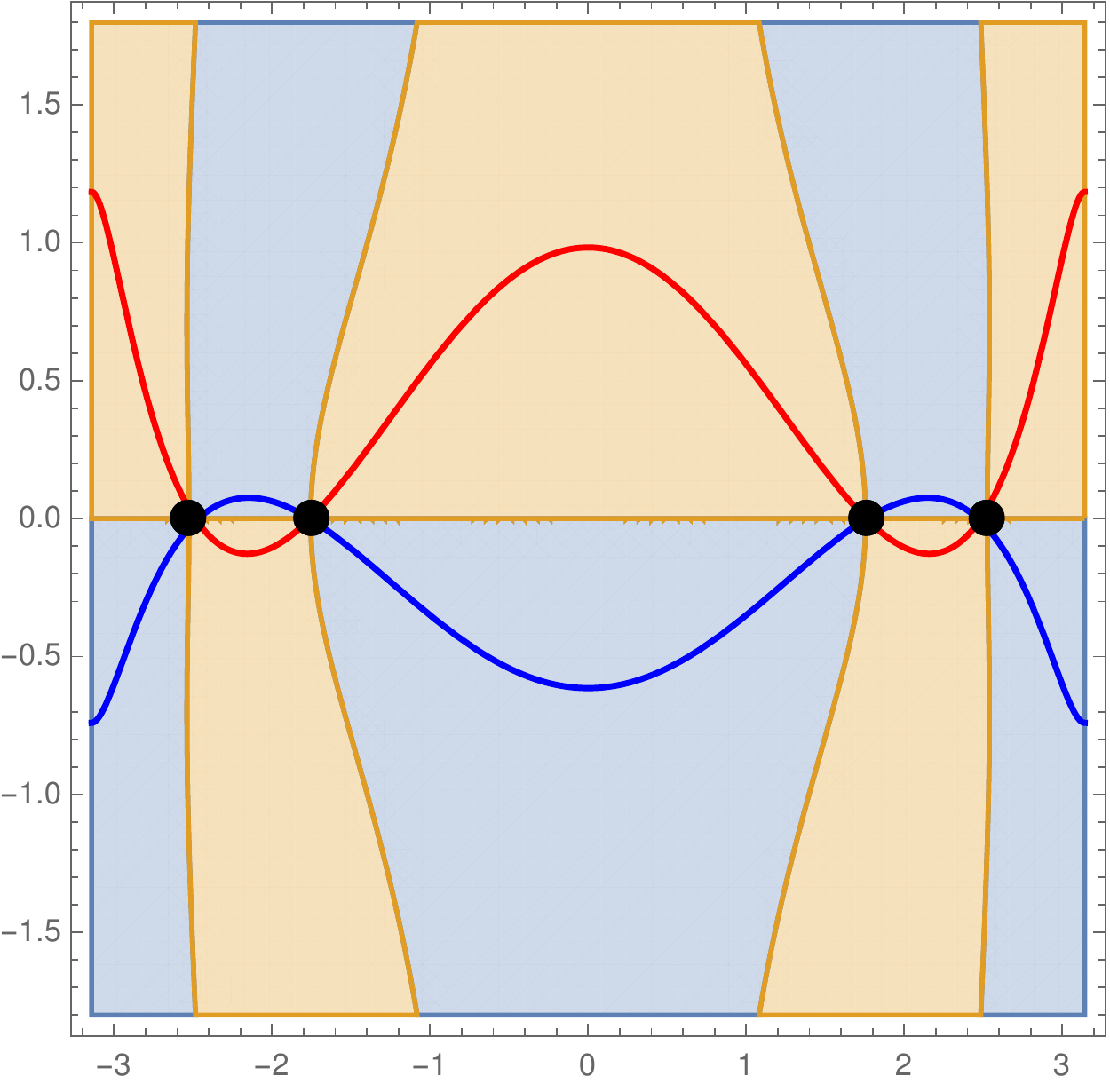}};
 \node at (13.8,0) {\includegraphics[width=0.24\textwidth]{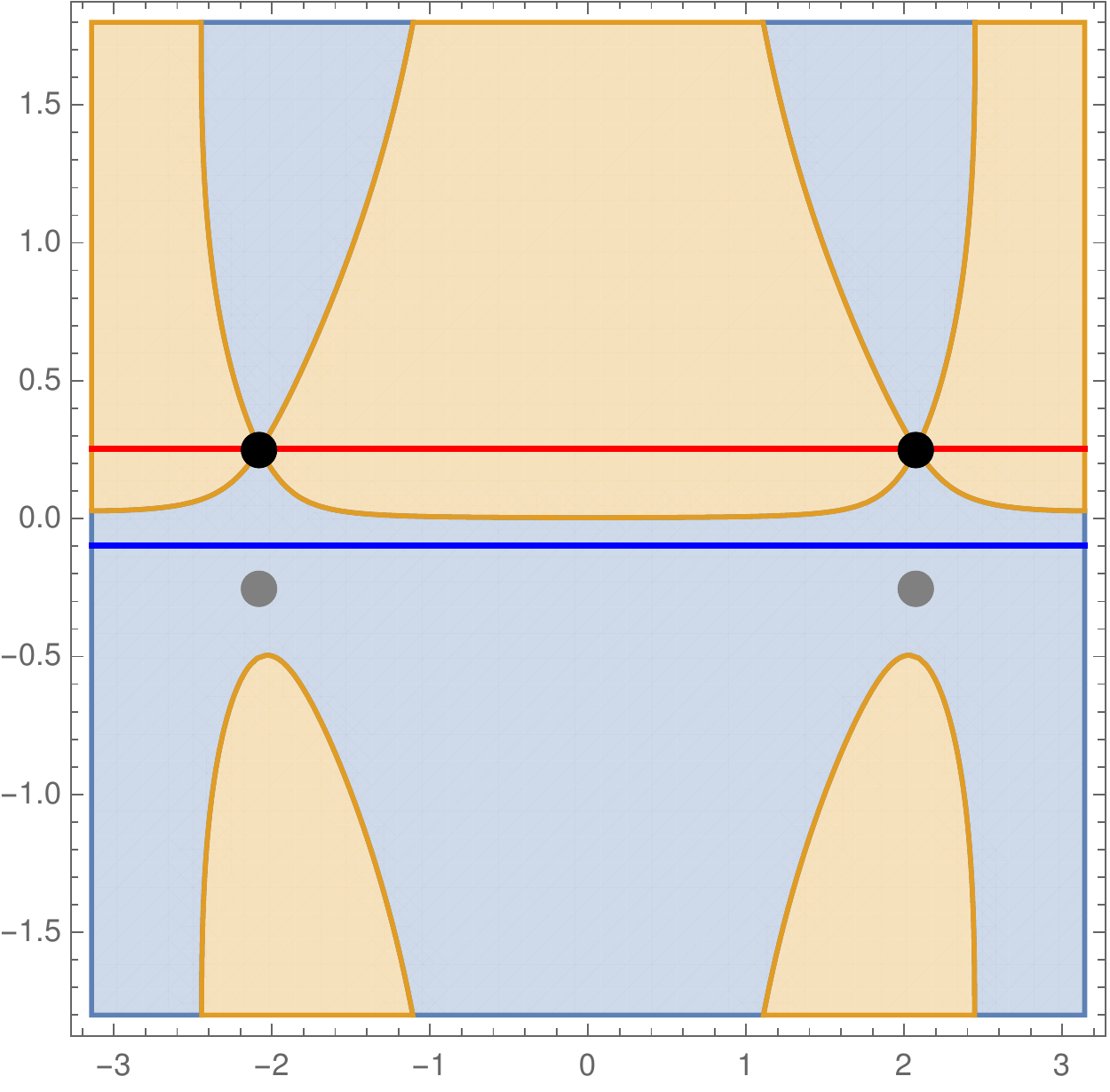}};
 \draw (0,2.5) node {(a) $\rho(X,Y)=0$};
 \draw (4.6,2.5) node {(b)\quad $0<\rho(X,Y)<1$};
 \draw (9.2,2.5) node {(c)\quad $0<\rho(X,Y)<1$};
 \draw (13.8,2.5) node {(d)\quad $\rho(X,Y)=1$};
 \end{tikzpicture}
 \caption{Possible contours at $Y=0$, for several values of $X$. (a): $X=1.6$, frozen region with density zero. (b) $X=0$, Fermi sea region. The density is $\frac{1}{2\pi}$ times the distance between the two real saddle points. (c) $X=-0.65$, split Fermi sea region. The density is $\frac{1}{2\pi}$ times the distance between the two real saddle points in the middle plus the distance from $-\pi$ to the leftmost saddle point plus the distance from the rightmost saddle point to $\pi$. (d) $X=-0.75$, frozen region with density one.}
 \label{fig:alphacontour_y0}
\end{figure}

\begin{figure}[htbp]
 \includegraphics[width=0.325\textwidth]{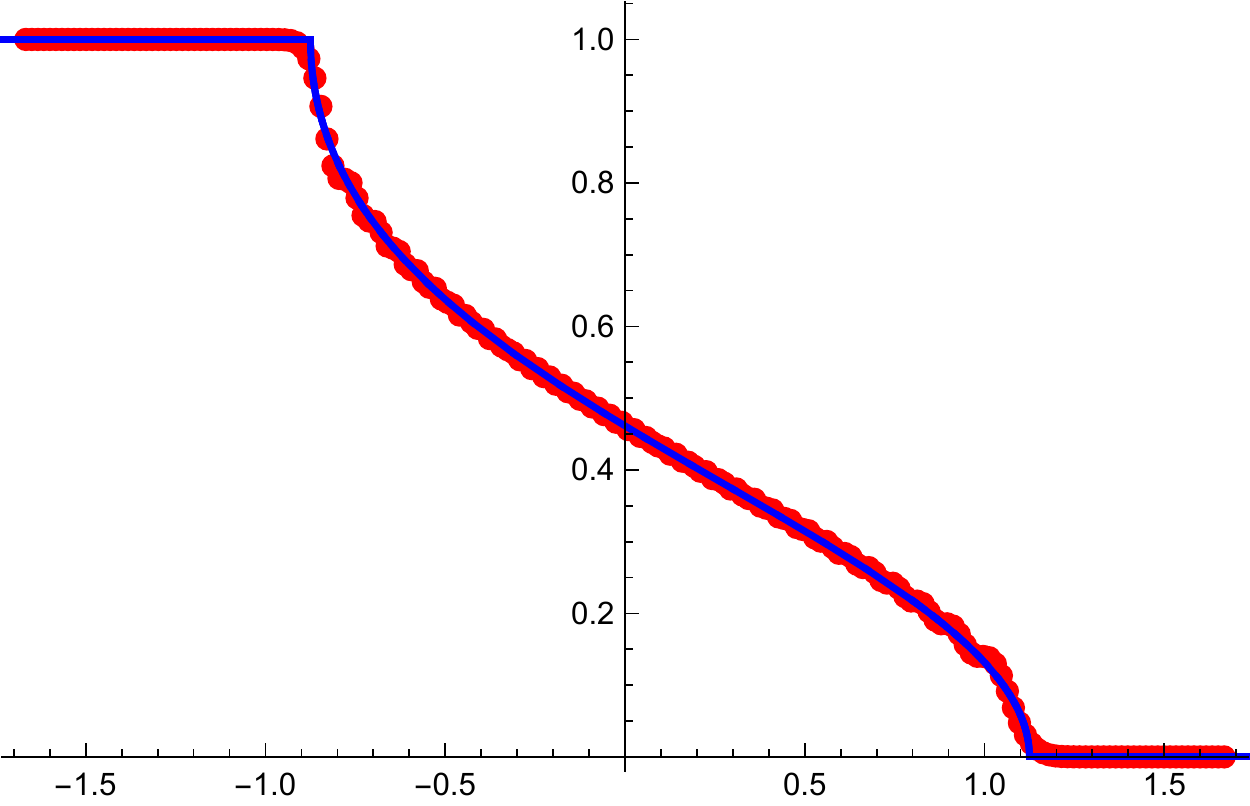}
  \includegraphics[width=0.325\textwidth]{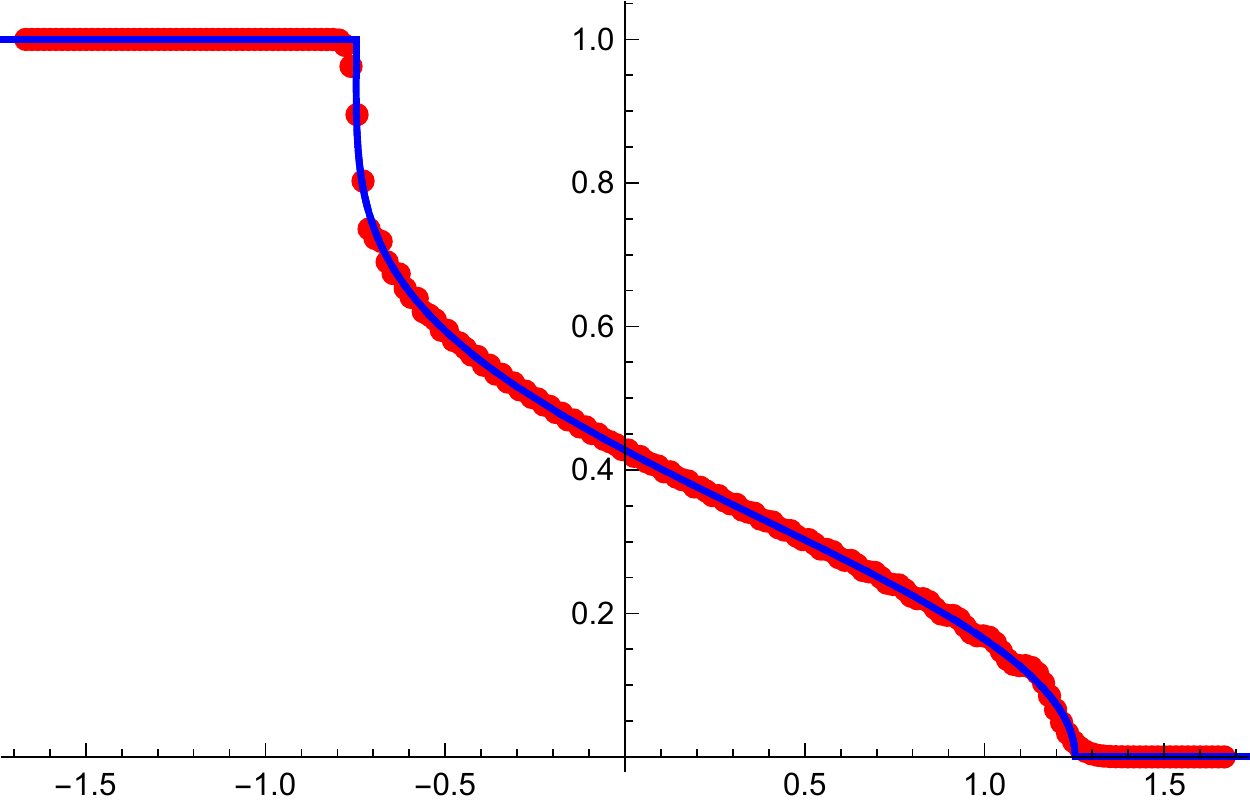}
   \includegraphics[width=0.325\textwidth]{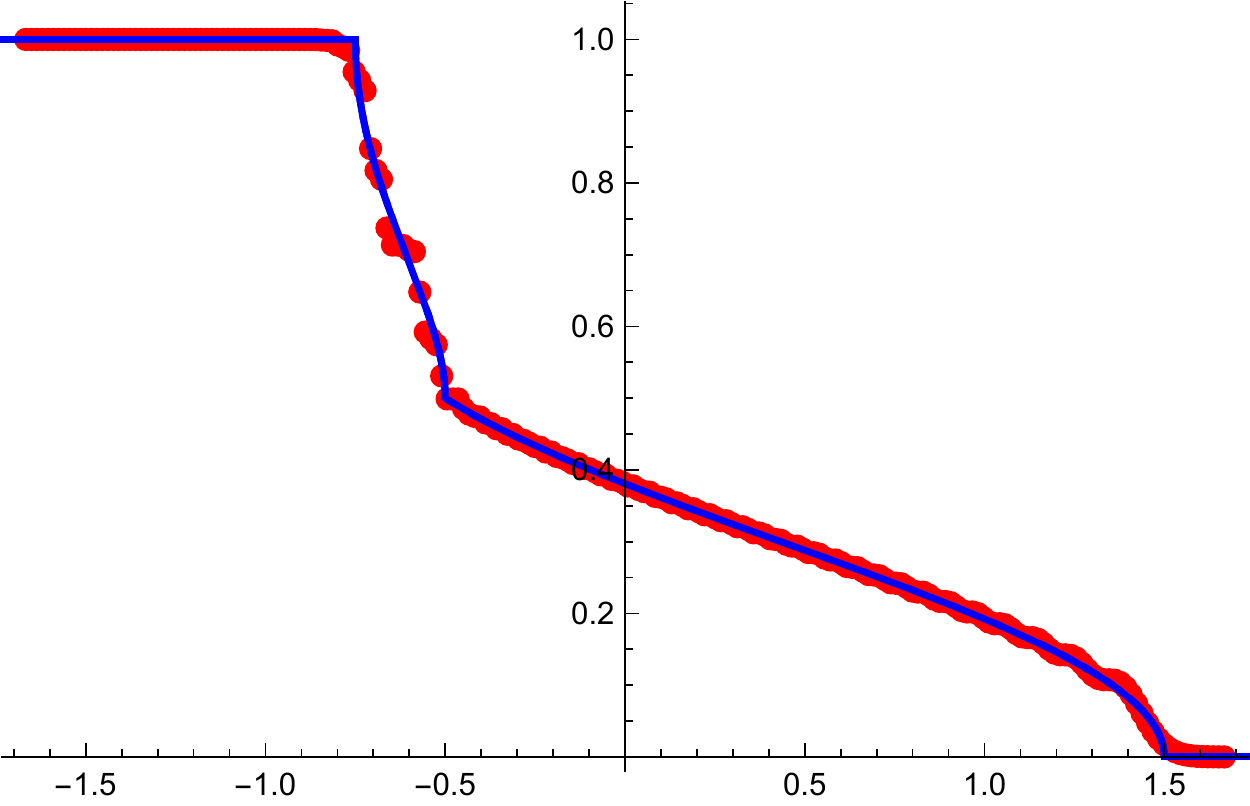}
   \caption{Density profile $\rho(X,Y=0)$ for $\alpha=1/16$ (top left), $\alpha=1/8$ (top right), and $\alpha=1/4$ (bottom). The blue curves are the analytical formulas (\ref{eq:density_1}),(\ref{eq:density_2}), the red dots numerical checks using a finite but large system with length $L=200$ and $R=60$.}
   \label{fig:densityplots_y0}
\end{figure}

This short section is devoted to the density profile of the time slice $Y=0$, which is always well-behaved. In that specific case, $\varphi(k)=-\ci (kX-\tilde{\varepsilon}(k))$ is pure imaginary for real $k$. It is possible to use the contour deformation method of the previous section to get the density profile, but the method of Ref.~\cite{real_time} allows for a more compact expression. Making the change of variable $K=\frac{k+q}{2}$ and $Q=k-q$, and linearizing the integral over $Q$ yields
\begin{equation}
 \rho(X,Y)=\int_{-\pi}^{\pi} \frac{dK}{2\pi}\Theta(-X+\tilde{\varepsilon}'(K)).
\end{equation}
where 
\begin{equation}
 \tilde{\varepsilon}'(K)=\cos K+2\alpha \cos 2K,
\end{equation}
and $\Theta$ is the Heaviside step function. 
This density coincides with the ground state density profile corresponding to a dispersion $\tilde{\varepsilon}'$, which has been studied e.g. in Ref.~\cite{Stephan_edge}. For $\alpha<1/8$ there can be either zero (in which case $\rho=0$ or $\rho=1$) or two real solutions $z_1$, $z_2=-z_1$ to the equation $X=\tilde{\varepsilon}'(K)$, which leads to a Fermi-sea ground state
\begin{equation}
\rho(X,Y)= \frac{z_2-z_1}{2\pi}.
\end{equation}
For $\alpha>1/8$ an interesting feature is worth noting: for some values of $X$, there can also be $4$ real solutions $z_1$, $z_2$, $z_3=-z_2$, $z_4=-z_1$, in which case the density reads
\begin{equation}
 \rho(X,Y)=\frac{(z_1+\pi)+(z_3-z_2)+(\pi-z_4)}{2\pi}
\end{equation}
This has the structure of a split Fermi sea \cite{split_fs}. The result can be recovered from the saddle point method, as shown in figure \ref{fig:alphacontour_y0}. 
Interestingly, the expressions for the saddle points simplify greatly, and can be solved in terms of two quadratic equations. This is obvious in the trigonometric version, while in terms of (\ref{eq:spe_full}) we get a so-called palindromic quartic equation. Summarizing, the density profile is given by
\begin{equation}\label{eq:density_1}
 \rho(X,0)=\left\{
 \begin{array}{ccc}
  1&,&X\leq -1+2\alpha\\\\
  \frac{1}{\pi}\arccos \left(\frac{-1+\sqrt{1+32\alpha^2+16\alpha X}}{8\alpha}\right) &,&-1+2\alpha\leq X \leq 1+2\alpha\\\\
  0&,& 1+2\alpha\leq X
 \end{array}
 \right.
\end{equation}
for $\alpha\leq 1/8$, and 
\begin{equation}\label{eq:density_2}
 \rho(X,0)=\left\{
 \begin{array}{ccc}
  1&,& X\leq -\frac{1+32\alpha^2}{16\alpha}\\\\
   1+\frac{1}{\pi}\arccos\left(\frac{-1+\sqrt{1+32\alpha^2+16\alpha X}}{8\alpha}\right)-\frac{1}{\pi}\arccos\left(\frac{-1-\sqrt{1+32\alpha^2+16\alpha X}}{8\alpha}\right)&,&-\frac{1+32\alpha^2}{16\alpha}\leq X\leq -1+2\alpha\\\\
   \frac{1}{\pi}\arccos\left(\frac{-1+\sqrt{1+32\alpha^2+16\alpha X}}{8\alpha}\right)&,&-1+2\alpha\leq X \leq 1+2\alpha\\\\
   0&,&1+2\alpha\leq X
 \end{array}
 \right.
\end{equation}
for $\alpha\geq 1/8$. The profiles for $\alpha=1/16,1/8,1/4$ are shown in figure \ref{fig:densityplots_y0}.
The split Fermi sea corresponds to the second line in (\ref{eq:density_2}) and connects to the west frozen region with density $1$ in figure \ref{fig:densityplots_y0}(c). One can also check that this split Fermi sea regime with four real roots requires the left hand site of (\ref{eq:spe_full}) to be palindromic in $\omega$, so can only occur at $Y=0$.

Generically, the density goes to $0$ or $1$ with a square root behavior. Such vanishing of the density is well-known to be associated to $R^{1/3}$-Tracy-Widom \cite{TracyWidom_tw} behavior for the distribution of the rightmost fermion (leftmost hole), see e.g. \cite{Johansson2000}. This is not true in the limiting case $\alpha=1/8$, since the density goes to $1$ with a fourth root behavior (illustrated in figure \ref{fig:densityplots_y0}, middle). As argued in Ref.~\cite{Stephan_edge}, the fluctuations of the leftmost hole are now on a scale $R^{1/5}$, and should converge to the higher order distribution studied in \cite{higher1,higher2,higher3,Multi}.  We refer to \cite{WBB} for a full saddle point analysis.

\section{The lower density case}
\label{sec:nw}
The main difference with domain wall boundary conditions is obtaining the Wiener-Hopf decomposition of $\varepsilon_n$, which is less explicit for $n> 2$. We discuss in detail the $n=2$ case first, which we dub ``N\'eel wall'', in section \ref{sec:neelwall}. The $n> 2$ treatment is similar, and postponed to section \ref{sec:generalsub}.  
\subsection{N\'eel wall}
\label{sec:neelwall}
For the N\'eel wall boundary conditions, an extra effective dispersion enters in the contour formula (\ref{eq:main_exact}). It can be written as
\begin{equation}\label{eq:somethg}
 \varepsilon_2(k)=\alpha \cos k+e_2(k)\qquad,\qquad e_2(k)=\frac{1}{2R}\log \cosh \left(2R\cos \frac{k}{2}\right).
\end{equation}
Therefore, we obtain
\begin{equation}
 \label{eq:omega2}
 \Omega_2(k)=R\left[\cos k-e_2(2k)\right],
\end{equation}
and
\begin{equation}
 \Phi_2(k)=-\ci k x-y\varepsilon(k)+\ci R\left[\alpha \sin 2k+\tilde{e}_2(2k)\right].
\end{equation}
There are two extra difficulties compared to domain wall. First, $\Omega_2$ is nonzero, and second, the evaluation of the Hilbert transform of $e_2$ is not straightforward for finite $R$.

Let us focus on $\Omega_2$ first. For large $R$, observe that $\Omega_2(k)\sim \frac{1}{2}\log 2$ for $\re k\in (-\pi/2,\pi/2)$, while $\Omega_2(k)$ has real part going to $-\infty$ for $\pi/2< |\re k|\leq \pi$. Since the integrand in the contour integral comes with a factor $e^{\Omega(k)+\Omega(k')}$, the effect of this term is to suppress contributions from the region outside the half-strip $\re k\in (-\pi/2,\pi/2)$. The real part of $k$ is directly related to the density, so densities larger than $1/2$ will be exponentially disfavored, consistent with intuition. For small $R$ this is not so, and memory of the microscopic staggering in the initial state is kept.

Second, it is possible to evaluate the Hilbert transform in the limit $R\to\infty$. Essentially, it coincides with the Hilbert transform of the $2\pi-$periodic extension of the function  $k\mapsto \cos \frac{k}{2}$, see (\ref{eq:somethg}). Computing explicitly all Fourier coefficients, we obtain
\begin{align}
 \tilde{e}_2(k)&\sim \sum_{p=1}^{\infty}\frac{(-1)^{p+1}}{\pi(p^2-1/4)}\sin (pk)\\
 &=\frac{2}{\pi}\cos \frac{k}{2}\textrm{arctanh}\left(\sin \frac{k}{2}\right)
\end{align}
to the leading order in $1/R$, for $k\in(-\pi,\pi)$. It is important to realize this asymptotic form of the Hilbert transform is not differentiable at $k=\pi$. This means at $k=\pi$ a finer analysis is in principle required, leading to $\tilde{e}'_2(\pi)=-R$, and some non trivial scaling function in the neighborhood of $\pi$. However, we won't need any of this in the limit $R\to \infty$ for fixed $X,Y$ under study here.
Hence to the leading order 
\begin{equation}
 \Phi(k)=-\ci k x-y\left(\cos k+\alpha\cos 2k\right)+\ci R\left(\alpha \sin 2k+\frac{2}{\pi}\cos k \,\textrm{arctanh}(\sin k)\right),
\end{equation}
and the saddle point equation for $\Phi$ reads
\begin{equation}
 \label{eq:neel_spe}
 X+\ci Y\left(\sin k+2\alpha \sin 2k\right)-\left(2\alpha\cos 2k+\frac{2}{\pi}\left[1-\sin k \,\textrm{arctanh}(\sin k)\right]\right)=0.
\end{equation}
This is a transcendental equation, but the solutions have a simple structure in the strip $\re k \in(-\pi/2,\pi/2)$, for any $\alpha\geq 0$. Depending on $X$ and $Y$, we observe that those are either $2$ pure imaginary numbers, or come in a \emph{single} pair of anticonjugated complex numbers $z,-z^*$ (we take $\re z\in [0,\pi/2)$). While we unfortunately we do not have a formal proof of these statements, we checked them numerically for a wide range of values for $\alpha,X,Y$. 

\begin{figure}[htbp]
 \begin{tikzpicture}
 \node at (0,0) {\includegraphics[width=0.25\textwidth]{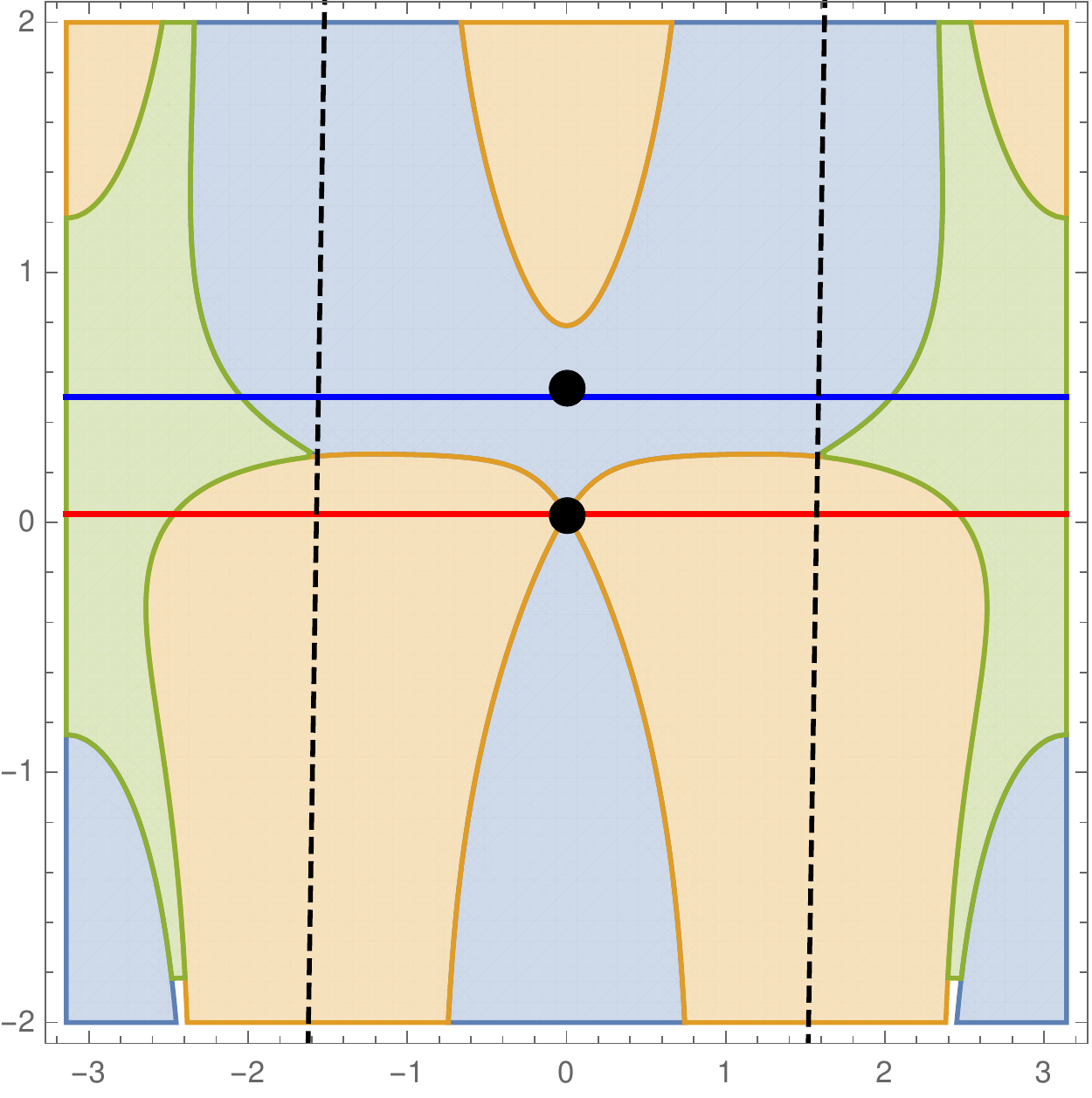}};
 \node at (6.5,0) {\includegraphics[width=0.24\textwidth]{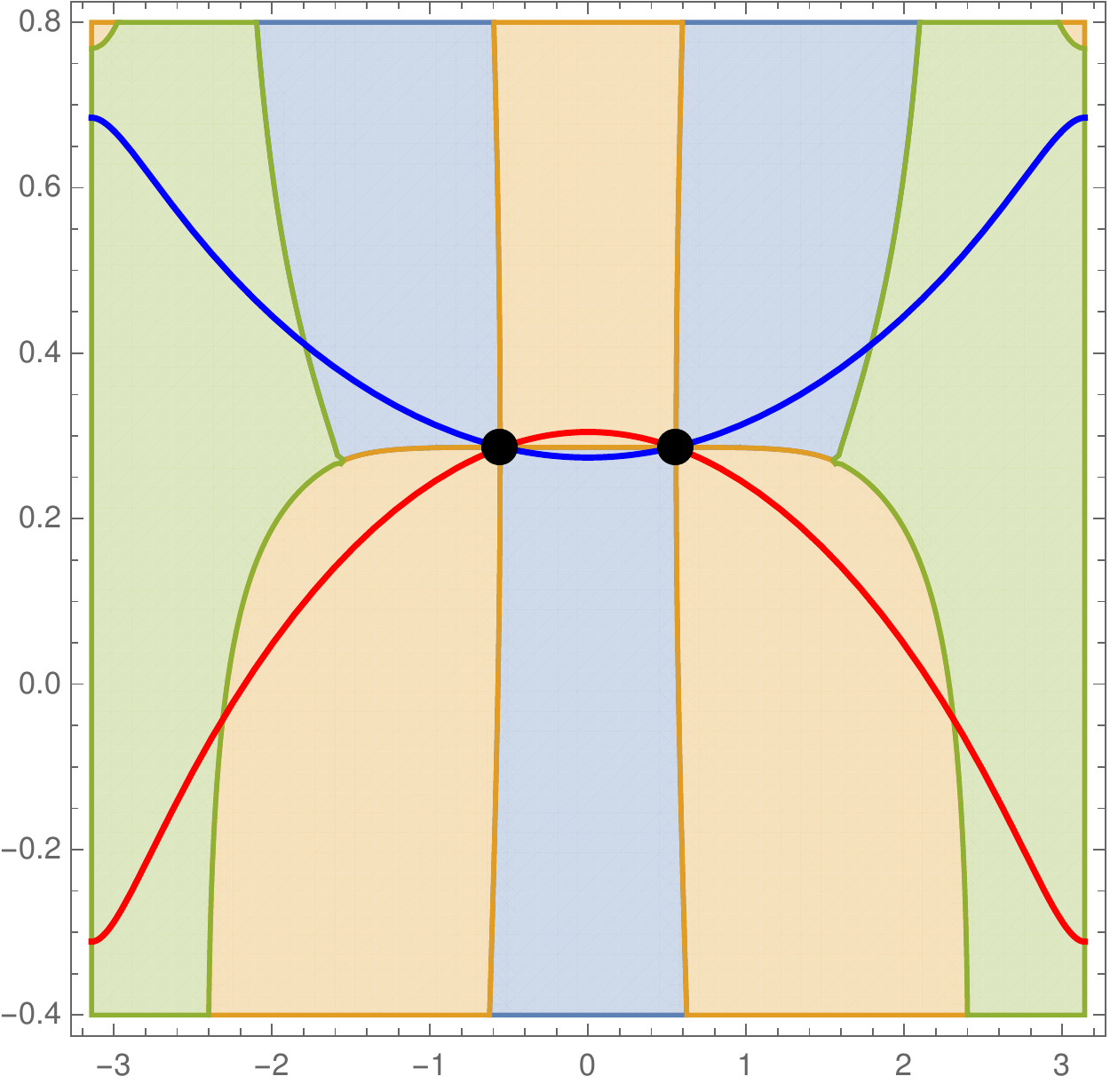}
 };
 \node at (13,0) {\includegraphics[width=0.24\textwidth]{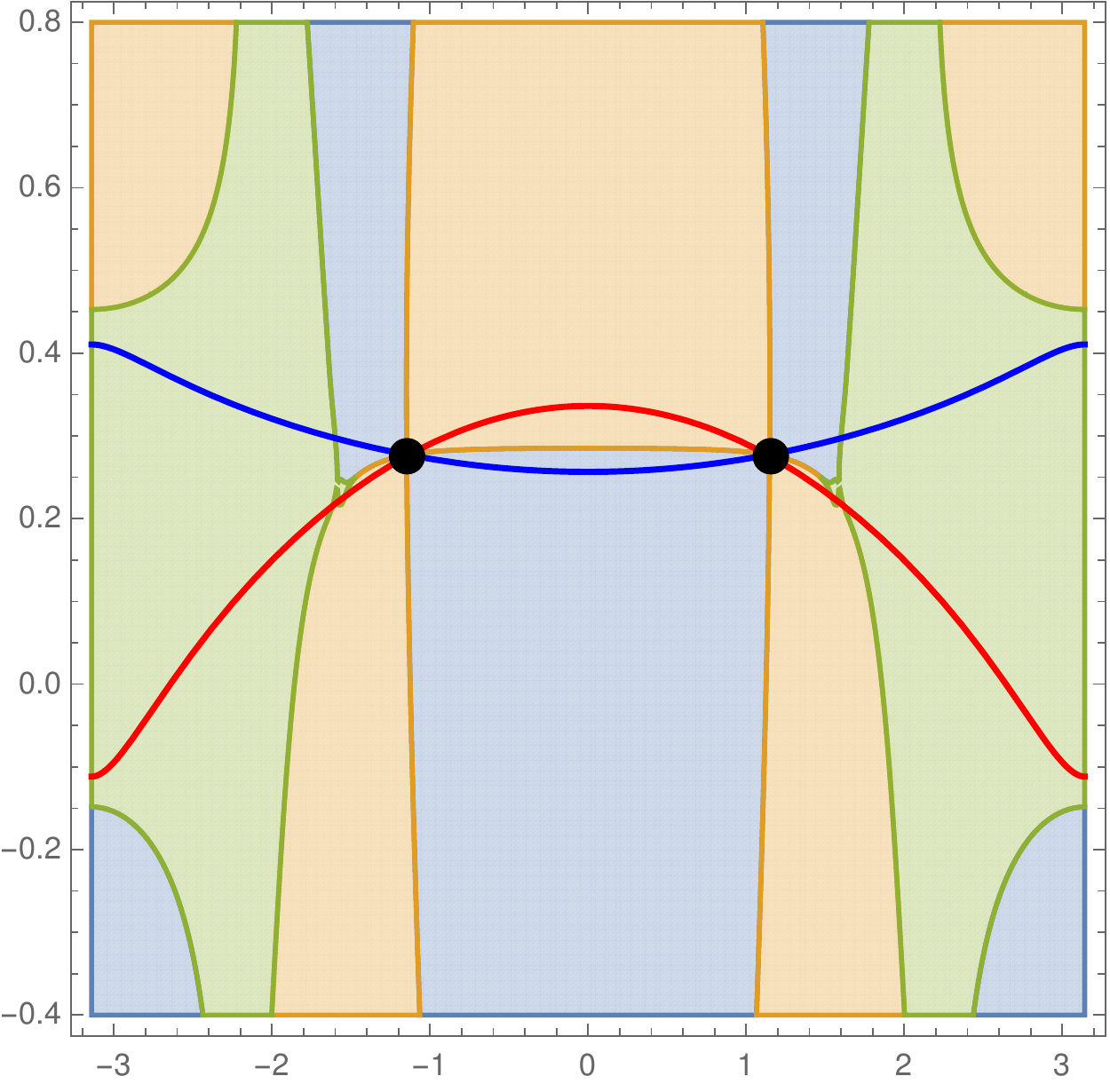}
 };
 \draw (0,2.5) node {(a) $\rho(X,Y)=0$};
 \draw (6.5,2.5) node {(b)\quad $0<\rho(X,Y)<1$};
 \draw (13,2.5) node {(c)\quad $0<\rho(X,Y)<1$};
 \end{tikzpicture}
 \caption{Possible integration contours for N\'eel wall and $\alpha=2$. In the blue region (C1) holds but not (C2), in the orange region (C2) holds but not (C1). Both conditions hold in the green region, which means it is available for both contours. Note $|\re k|\geq \pi/2$ in this region. $C_k$ is the thick red line, while $C_q$ is the thick blue line, with the saddle points represented by black dots. (a) A point $X=4.5$, $Y=-0.5$ in the east frozen region, with a pure imaginary saddle point. It is straightforward to find a deformation ensuring exponential decay to zero. (b) A point $X=1.9$, $Y=-0.5$ in the fluctuating region. A correct deformation can be found at the expense of having the two contours crossing on a curve joining the two saddle points $z$ and $-z^*$, leading to formula (\ref{eq:nw_z}). (c) A similar point $X=-2.5$, $Y=-0.5$ with a higher density.}
 \label{fig:nw_contours}
\end{figure}
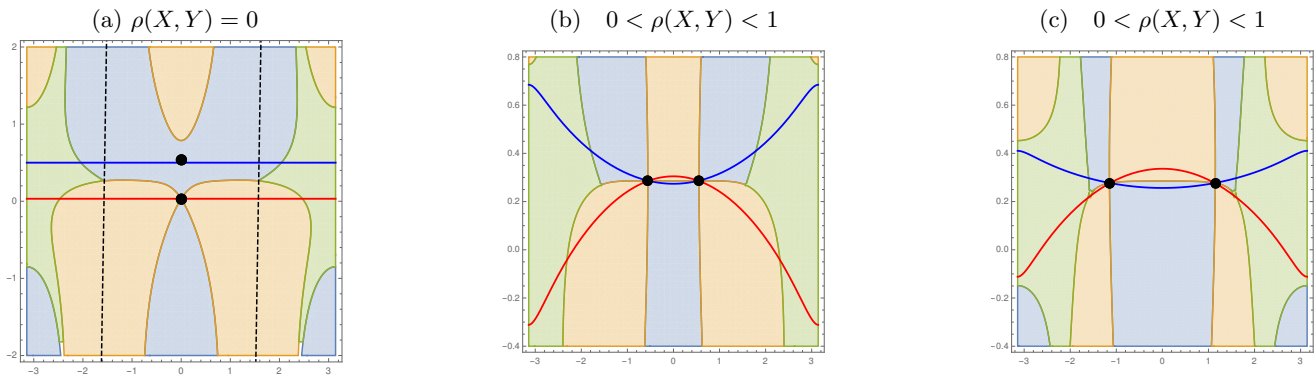

With this at hand, it is not difficult to anticipate the right deformation. We make the contour go through the saddle points as we did in the previous section \ref{sec:nn}. The region where the saddle points are pure imaginary corresponds to the frozen part of the profile, with density $0$, while the other region is the fluctuating region. Since the contribution stemming from $e^{\Omega_2(k)+\Omega_2(k')}\sim 2$ is exactly canceled by the residue of $\frac{1}{z^2-1}$ at $z=1$, we get a density
\begin{equation}\label{eq:nw_z}
 \rho(X,Y)=\frac{\re z}{\pi}.
\end{equation}
Importantly, those are the only saddle points that matter, and the deformation can always be performed to get exponential decay (up to a residue term which gives precisely the density, as before). More precisely, denote by (C1) the condition $\textrm{Re}[\Phi(k)-\Omega(k)]>\textrm{Re}[\Phi(z)-\Omega(z)]$ and (C2) the condition $\textrm{Re}[\Phi(k)+\Omega(k)]<\textrm{Re}[\Phi(z)+\Omega(z)]$. Then, if one can perform a deformation such that (C1) holds for $C_q$ and (C2) holds for $C_k$ -- possibly up to a finite number of points, the integrand is exponentially small almost everywhere and the double integral decays to zero.
The relevant integration contours and regions are shown in figure \ref{fig:nw_contours}.

Therefore, for any $\alpha\geq 0$ there are no crazy regions, the density is well defined $\in [0,\pi/2)$ for any $X,Y$. Even though the solution $z$ is given in implicit form, a number of limiting cases can be worked out explicitly. This includes the arctic curve, as well as the $X\to -\infty$ limit.
\paragraph{The arctic curve and edge behavior.}
This can be extracted from (\ref{eq:neel_spe}), by setting $k=\epsilon+\ci\,\textrm{arcsinh}\,s$, expanding for small $\epsilon$, and asking that both real and imaginary parts be zero. We obtain this way the curve in an explicit but complicated form. The expression simplifies a lot in the particular case $\alpha=0$, which we report below:
\begin{align}
 X_{\rm a}(s)&=\frac{2}{\pi(1+s^2)},\\
 Y_{\rm a}(s)&=\frac{2}{\pi}\left(\frac{s}{1+s^2}+\arctan s\right),
 \end{align}
for $s\in \mathbb{R}$. This can be also rewritten as
\begin{equation}
 \pm Y_{\rm a}= \sqrt{X_{\rm a}\left(\frac{2}{\pi}-X_{\rm a}\right)}+\frac{1}{\pi}\arccos \left(\pi X_{\rm a}-1\right).
\end{equation}
By a similar argument, one can establish that the density always vanishes as a square root in the vicinity of the arctic curve. For example at $Y=0$ this simply follows from the expansion of the saddle point equation
\begin{equation}
 X-X_{\rm a}+\left(\frac{2}{\pi}+4\alpha\right) k^2=0
\end{equation}
about the arctic curve $X_{\rm a}=\frac{2}{\pi}+2\alpha$. For $\alpha\geq 0$ considered in this paper, the coefficient of $k^2$ cannot vanish, 
meaning the distribution of the rightmost fermion should always converge to the Tracy-Widom distribution. 
\paragraph{$X\to-\infty$ asymptotics.} Another feature of N\'eel wall initial states is that there is no sharp arctic curve on the west side. This is due to the singularity of the Hilbert transform $\tilde{e}_2(k)$ at $k=\pi$ which translates into a logarithmic divergence of $\frac{d\Phi}{dk}$ at $k=\pi/2$, contrary to standard situations which give rise to sharp arctic curves as happens e. g. on the east side. Expanding the saddle point equation in the vicinity of $k=\pi/2^-$ leads to 
\begin{equation}\label{eq:expodecay}
 1/2-\rho(X,Y)\sim \frac{2}{\pi}e^{\pi \alpha-1}e^{\pi X/2}\cos \left(\frac{\pi Y}{2}\right) 
\end{equation}
for $Y\in (-1,1)$ in the limit $X\to -\infty$. Recall that this result is only valid in the limit $R\to \infty$ first, and then $X\to -\infty$. There is another asymptotic regime associated to $X$ of order $R$, in which case the singularity in the Hilbert transform is regularized, and decay to density $1/2$ is even faster.

\begin{figure}[htbp]
\includegraphics[width=0.6\textwidth]{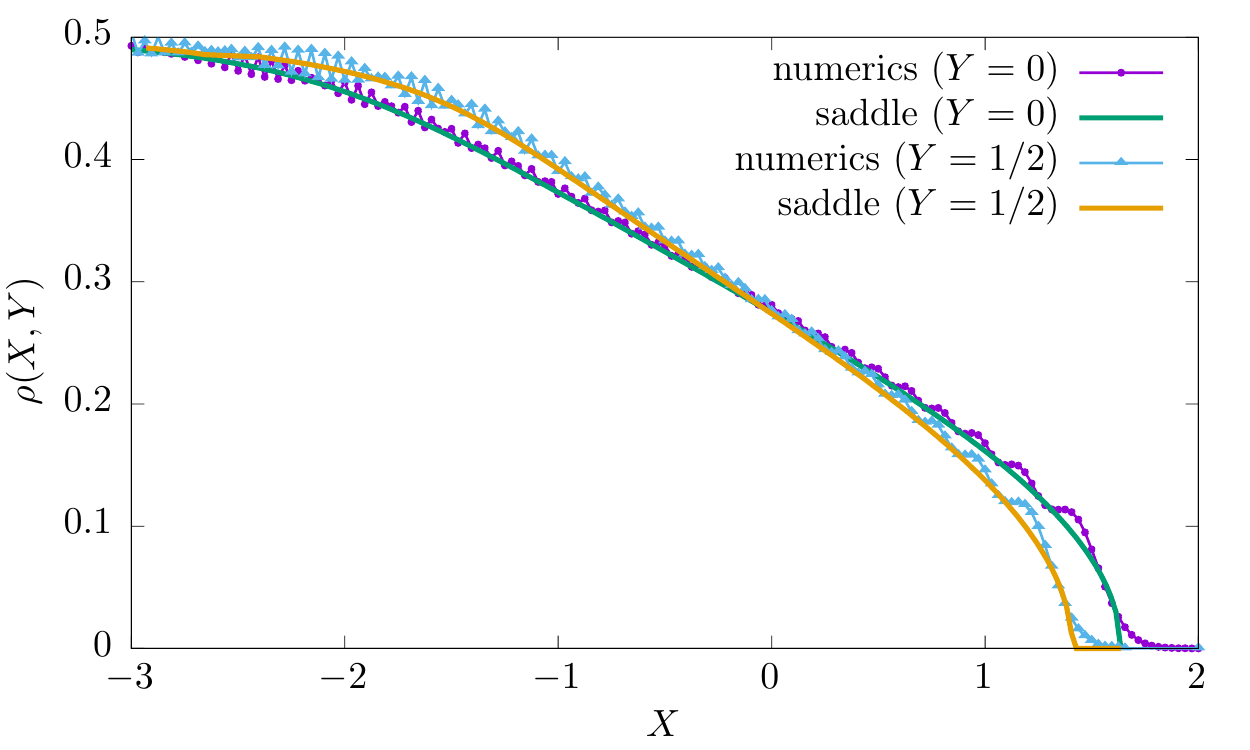}
 \caption{Numerical checks of formulas (\ref{eq:neel_spe}),(\ref{eq:nw_z}), for $\alpha=1/2$, at both $Y=0$, $Y=1/2$. The yellow and green curves correspond to analytic results obtained from solving the saddle point equation (\ref{eq:neel_spe}). The violet dotted and the blue triangle lines are the numerical data for $R=32$, and show very good agreement. Observe once again the absence of a sharp arctic curve to the left, replaced instead by the exponential decay (\ref{eq:expodecay}).}
 \label{fig:checks}
\end{figure}

\subsection{General $n$}
\label{sec:generalsub}
The general $n$ case is very similar to $n=2$, we provide some explicit formulas in this section for completeness. The effect of the potential $\Omega_n(k)$ is to effectively reduce the size of the Brillouin zone, since for large $R$ we have $\Omega_n(k)\sim \frac{1}{2}\log n$ for $k\in (-\pi/n,\pi/n)$, and otherwise its real part goes to $-\infty$.

The next step is to compute the Hilbert transform of (the $2\pi$-periodic extension of) $f_d:k\mapsto \cos d k$ for $d\in (0,1)$, which is given by
\begin{equation}\label{eq:general_hilbert}
 \tilde{f}_{d}(k)= \frac{\sin \pi d}{\pi d}\int_{0}^{\infty}du \frac{\sinh u \,\sin k}{\cosh (u/d)+\cos k}.
\end{equation}
This can also be expressed in terms of a hypergeometric function. 
Setting $d=1/n$ and $d=2/n$ in the previous equation, we obtain
\begin{align}
 \tilde{\varepsilon}_n(k)
 &=\frac{n}{\pi}\sin \frac{\pi}{n}\sin k
 \int_0^\infty du \sinh u\left[\frac{1}{\cosh nu+\cos k}+\frac{\alpha \cos \frac{\pi}{n}}{\cosh \frac{nu}{2}+\cos k}\right],
\end{align}
where the cases $n=1$ and $n=2$ are understood as limits. This leads to the saddle point equation 
\begin{equation}\label{eq:mostgen_spe}
 X+\ci Y \left(\sin k+2\alpha \sin 2k\right)-
 \frac{n^2}{\pi}\sin \frac{\pi}{n}
 \int_0^\infty du \sinh u \left[
 \frac{1+\cos nk \cosh nu}{(\cosh nu+\cos nk)^2}+\alpha\frac{1+\cos nk \cosh \frac{nu}{2}}{(\cosh \frac{nu}{2}+\cos nk)^2}
 \right]\;=\;0.
\end{equation}
We again observe that there are either two pure imaginary solutions, or two complex anticonjugated solutions $z,-z^*$ in the strip $\re k\in(-\pi/n,\pi/n)$. The limit shape follows from this, since once again contributions outside of this region are exponentially suppressed due to $\Omega_n(k)$. Hence we get a density $\rho(X,Y)\sim \frac{\re z}{\pi}\in [0,1]$. Few features can be extracted explicitly from this equation, however, the exponential decay for $X\to-\infty$ can still be checked.

A conclusion from this study is that for any $n\geq 2$, the density is well-behaved in the scaling limit. This is consistent with the intuition that lowering the density helps get rid of potentially proliferating minus signs in the probability distribution (\ref{eq:nonproba_dist}). 

\section{Discussion and hydrodynamics}
Let us give a short summary of our findings. We have considered a free fermionic model in imaginary time, also known as the XX chain (after Jordan-Wigner transformation). This model is known to give rise to limit shapes, related to the polynuclear growth model \cite{Spohn_prl}. We then perturbed it by adding a next nearest neighbor hopping term. This perturbation is not always positive, due to the fermionic anticommutation relations. The nonpositive part of the perturbation is typically either exponentially suppressed in the thermodynamic limit, or creates new crazy regions in which the density is not between 0 and 1 anymore. \\For initial states with a sufficiently low density, the crazy regions completely disappear, and the model behaves as if it were positive, no matter how strong the amplitude of the perturbation, set by $\alpha$, is. We demonstrated this by performing a saddle point analysis of the exact formula (\ref{eq:main_exact}), and careful numerical checks. Along the horizontal line, $Y=0$ in the middle, the model is always positive due to symmetry considerations, and new limit shape results were obtained as a byproduct of our analysis.

The (intuitively natural) competition between the range of the hoppings and the density can also be investigated by looking at other dispersions, for which generalization of our results, in particular through (\ref{eq:general_hilbert}), is straightforward. For example, we checked that with a dispersion of the form 
$\varepsilon(k)=\cos k+\beta \cos (4k)$ with $\beta>0$, an initial state with density $d=1/2$ on the west ($n=2$) does not prevent the appearance of crazy regions. However, $d\leq 1/4$ does, once again through a mechanism of reducing the number of relevant saddle points to $2$.

While our results were based on exact free fermions lattice techniques, there are also softer variational \cite{Nienhuis_1984,variationaldimers,kenyon2007} or hydrodynamic \cite{Abanov_hydro} approaches to limit shapes. In the language of one-band fermions discussed in the present paper, the hydrodynamic equations are given by some generalized complex Burgers equation 
and their solution can be parameterized as \cite{Abanov_hydro}
\begin{equation}
 R \,G(z)=x+\ci y \varepsilon'(z),
\end{equation}
for some analytic function $G$, which encodes boundary conditions. Given a function $G$, it is easy to check whether the boundary conditions are satisfied, but the inverse problem, given a boundary condition, is in general tremendously difficult. This is of course the main bottleneck to applying this method. The case of ``emptiness'' boundary conditions --where a segment does not contain any fermion-- has been worked out in the original paper, while the solution for domain wall is given in \cite{Stephan_ebc}. Of course in the latter case, it is easy to make the correct guess by reverse-engineering the exact saddle point result, and check that this indeed solves the hydrodynamic equation. \\By the same logic, the SPE (\ref{eq:spe_full}) allows us to solve the hydrodynamic problem for boundary conditions with density $d\in (0,1]$ to left, and zero to the right. Considering only the probabilistic case $\alpha=0$ for simplicity, the correct hydrodynamic equation reads for large $R$
\begin{equation}\label{eq:mostgeneral_spe}
x+\ci y \sin (d\kappa)-R\tilde{v}_d(\kappa)=0,
\end{equation}
where $\tilde{v}_d(k)$ is the Hilbert transform of $v_d:k\mapsto \sin (dk)$ extended to a $2\pi$ periodic function, which can be computed from the derivative of  (\ref{eq:general_hilbert}).
This coincides with (\ref{eq:mostgen_spe}) when $1/d=n$ is an integer, after setting $z=d\kappa$. In the fluctuating region $\kappa$ is unique provided we impose $\re \kappa \in [0,\pi]$, and the density is now \footnote{Alternatively, one could consider the hydrodynamic equation $x+\ci y \sin z-R\tilde{v}_d(z/d)=0$ with density $\frac{z-z^*}{2\pi}$.} given by $d\frac{\kappa-\kappa^*}{2\pi}=\frac{z-z^*}{2\pi}$. One can check from the definition of the Hilbert transform that the required boundary conditions are indeed satisfied, that is, solutions are of the form $\kappa=\pi+\ci \beta$, leading to a density $d$, for any $x<0$ and $y=\pm R$. Similarly $\kappa$ is pure imaginary, leading to zero density, for $x>0$ and $y=\pm R$. 
 Hence generalization to any $d$ can be easily achieved with hydrodynamic arguments, even though obtaining this from the saddle point method would require more work \footnote{Already $d=2/3$ would require the inversion of a block-Toeplitz matrix, which can be done but is more difficult.}, perhaps in the spirit of what was done in Refs.~\cite{Boutillier2012,Mkrtchyan2011} for a simpler model. Let us emphasize once again that for $d\neq 1$, the singularity of $v_d$ at $k=\pi$ has important consequences; in particular it implies the absence of an arctic curve on the west part, replaced by a slower exponential decay.


The real time evolution from the initial state $\ket{\psi_n}$ can be recovered from the exact propagator, simply by performing a Wick rotation $y=\ci t$ and $R\to 0^+$. In this case, there are no subtleties associated to effective dispersions and Hilbert transforms. In particular, the associated hydrodynamic equations follow from the stationary phase approximation (see e.g. \cite{real_time})
\begin{equation}\label{eq:spe_real}
 x-t\sin \kappa=0,
\end{equation}
where solutions in the fluctuating region are real and come also in pairs $\arcsin \frac{x}{t}$, $\pi-\arcsin \frac{x}{t}$, and lead to a density $\rho(x,t)=\frac{d}{\pi}\arccos \frac{x}{t}$. Importantly, the previous equation does not coincide with the Wick rotated equation (\ref{eq:mostgeneral_spe}), which gives a density $\rho(x,t)=\frac{1}{\pi}\arccos \frac{x}{t}$, unless $d=1$, which has already been subject to several studies \cite{antal1999transport,antal2008,EislerRacz,SabettaMisguich,real_time,Moriya_2019}. There is no contradiction, since (\ref{eq:mostgeneral_spe}) was established by first assuming $R$ to be large, and then sending $R\to 0^+$, which is not mathematically justified. \\This observation does not only illustrate the fact that the hydrodynamic limit and the Wick rotation do not commute in general, but has potential consequences regarding the quantum field theory treatment of entanglement growth of integrable systems in out of equilibrium setups. Indeed, the most convenient approach is to relate entanglement to expectation values of twist operators in euclidean conformal field theory, and then perform the Wick rotation \cite{calabrese2006time,calabrese2007entanglement}. This was later generalized \cite{dubail2017conformal} to a class of  inhomogeneous setups, which includes $\ket{\psi_1}$, yielding analytical predictions in perfect agreement with numerics. A second method, in which $R$ is seen as a small UV cutoff to sidestep the additional technical difficulties associated with imaginary time problems and Hilbert transforms was also put forward in the same paper (see \cite{QGHD,EE_gen} for even more general results). 
The conclusion from our study is that the first method, while more elegant, would probably yield incorrect result for $\ket{\psi_{n}}$ with $n\geq 2$, since Wick rotating the imaginary time hydrodynamic limit does not give the correct density. It would also be interesting to investigate whether the method also breaks down for ground states with a certain density $d$, which can be seen as a complicated superposition of real space product states such as the ones studied in the present paper. 

While we focused all our attention on the density here, correlation functions are of course interesting, and should also follow from a saddle point analysis of the exact formula (\ref{eq:main_exact}). The general wisdom is that long range correlations in the fluctuating region are described by a conformal field theory in curved space (\cite{kenyon2007,Allegra_2016,astala2020dimer}), with the conformal structure set by the limit shape. We also expect this to hold for any (noncrazy) fluctuating region, but it would be interesting to check that, and possibly use that to compute fluctuations of more complicated observables.

Finally, an obvious limitation of the present paper is that only free fermions models were considered. We nevertheless very much expect our phenomenology to extend to interacting systems. For example, the interacting XXZ spin chain with boundary condition $\ket{\psi_1}$ has been studied in Reference \cite{Return_Stephan}, and does not suffer from a sign problem. The same goes for next nearest neighbor spin interactions, such as $J_1-J_2$ chains. However, it should be possible to perturb the XXZ Hamiltonian by adding a term proportional to the second logarithmic derivative of the transfer matrix. This extra term would generalize our next nearest neighbor hopping term, and introduce non trivial signs in the problem, presumably leading to similar results. Note that an analytical study would be extremely challenging; in that case one would probably have to rely on numerical methods such as Monte Carlo, or, possibly better because of the sign problem, DMRG simulations.

\acknowledgments
We are grateful to J\'er\'emie Bouttier, Pasquale Calabrese, Filippo Colomo, J\'er\^ome Dubail, Fabio Franchini, Jacopo Viti, Harriet Walsh for several useful discussions. SB was partially supported by the ANR-18-CE40-0033 grant ('DIMERS').

\hfill
\newpage
\appendix
\section{Totally positive dispersions}
\label{app:totalpos}
We discuss in this short appendix which dispersions are guaranteed to provide a totally positive model --which implies non-negative Boltzmann weights-- generalizing the simple argument given in the introduction that this is the case for $\ve(k)=\cos k$. We consider a fermion chain on $\mathbb{Z}$ with arbitrary dispersion $\ve(k)$. The time evolved state is given by
\begin{equation}
 e^{\tau H}\ket{\psi}=\sum_{\mathcal{C}} \ac{\tau}\ket{\mathcal{C}}.
\end{equation}
We now ask that $a_\mathcal{C}(\tau)\geq 0$ for any $\tau\geq 0$, any initial state of the form 
\begin{equation}\label{eq:initpos}
 \ket{\psi}=c_{j_1}^\dag \ldots c_{j_N}^\dag\ket{0}
\end{equation}
 where the fermions are ordered, $j_1<\ldots<j_N$, and any state 
 \begin{equation}\label{eq:C}
\ket{\mathcal{C}}=c_{i_1}^\dag \ldots c_{i_N}^\dag \ket{0}
 \end{equation}
where the fermions are also ordered in the same way. The amplitude can be easily computed with Wick's theorem:
\begin{equation}
 a_\mathcal{C}(\tau)=\det_{1\leq a,b\leq N}\left(\braket{0|c_{i_a}e^{\tau H}c_{j_b}^\dag|0}\right),
\end{equation}
where
\begin{equation}
\braket{0|c_{i_a}e^{\tau H}c_{j_b}^\dag|0}=\int_{-\pi}^{\pi} \frac{dk}{2\pi}e^{-\ci k(i_a-j_b)}g(k) \qquad,\qquad g(k)=e^{\tau \ve(k)}. 
\end{equation}
Therefore, asking that all amplitudes be nonnegative amounts to asking that all determinants of finite minors of the doubly infinite Toeplitz matrix $(T_{ij})_{i,j\in \mathbb{Z}}$ with generating function --or symbol-- $g(k)$, $T_{ij}=\int_{-\pi}^{\pi} \frac{dk}{2\pi} e^{-\ci k(i-j)}g(k)$ be nonnegative. This property is called total nonnegativity or total positivity. The Edrei-Thoma theorem \cite{Aissen1952,Edrei,Thoma} classifies all possible symbols leading to totally nonnegative Toeplitz matrices (see also \cite{totalpos_review} for a review). The answer is that all totally nonnegative symbols may be written as
\begin{equation}\label{eq:totalpositiveform}
 g(k)=\exp\Big(\ci k p+A_0+A_1 e^{\ci k}+A_{-1}e^{-\ci k}\Big)\prod_{l=1}^{\infty} \frac{\left(1+B_l e^{\ci k}\right)\left(1+C_l e^{-\ci k}\right)}{\left(1-D_l e^{\ci k}\right)\left(1-E_l e^{-\ci k}\right)},
\end{equation}
where $p\in \mathbb{Z}$, $A_0\in [-\infty,\infty)$, all other coefficients are nonnegative, and the series $\sum_{l=1}^\infty(B_l+C_l+D_l+E_l)$ converges. Then, the choice $A_1=A_{-1}=\tau/2$ and setting all other coefficients to zero gives back $\ve(k)=\cos k$, which is indeed positive. However, the dispersion $\ve(k)=\cos k+\alpha \cos 2k$ studied in this paper is not of the form (\ref{eq:totalpositiveform}) for $\alpha>0$. Note also that the product part occurs for example while studying classical  dimers on the honeycomb lattice, which have $\ve(k)=\log([1+u e^{\ci k}][1+u e^{-\ci k}])$ (e.g. \cite{Allegra_2016}), in which case imaginary time needs to be an integer to ensure positivity, consistent with the discrete nature of the model both in the horizontal and vertical direction.

We finish with two remarks. First, it is of course possible to relax the total positivity constraint, and just ask for positivity within a subclass of initial states. A trivial example is that of single particle initial states $\ket{\psi}=c_i^\dag\ket{0}$, for which $\ve(k)=\sum_{p} a_p \cos (kp)$ is positive provided all $a_p\geq 0$. Second, since we are looking at distributions given by $\ac{R-y}\ac{R+y}$, the possible signs in both terms might cancel each other. This is the case for any dispersion, when $y=0$. However, other dispersions can lead to a positive model $\forall y$, such as $\ve(k)=-\cos k$, which is not of the form (\ref{eq:totalpositiveform}), but for which the sign of $\ac{\tau}$ can easily be shown not to depend on $\tau$. 

To summarize, there are several notions of positivity considered here, and restricting to states of the form (\ref{eq:initpos}):
\begin{itemize}
 \item Total positivity. $\ac{\tau}\geq 0$ $\forall \tau\geq 0$, $\ket{\psi}$, $\ket{\mathcal{C}}$. In a transfer matrix picture, this implies non-negative Boltzmann weights. 
 \item Partial positivity. $\ac{\tau}\geq 0$ $\forall \tau\geq 0$, $\ket{\mathcal{C}}$ and a given $\ket{\psi_0}$.
 \item Partial or total positivity of $\ac{R-y}\ac{R+y}$, depending on $y$. The case $y=0$ is always totally positive in that respect.
 \item Positivity of the density in the limit $R\to \infty$, with $x/R\in \mathbb{R}$ and $y/R\in(-1,1)$ fixed. An example is discussed in depth in section \ref{sec:nw}. It is not completely clear to us whether a stronger form of positivity holds in that case too.
\end{itemize}

\section{Semi-infinite Toeplitz matrices}
\label{app:toeplitz}
In this appendix, we gather some classical \cite{OPUC} results on semi-infinite Toeplitz matrices, and present exact formulae for their inverses, which follow from elementary linear algebra and Fourier analysis. Those will be necessary to derive the exact formula for the propagator (\ref{eq:main_exact}), which is done in appendix \ref{app:derivation}.

Consider a $2\pi$-periodic function $g(k)$, which we call symbol in the following. We write its Fourier coefficients as
\begin{equation}
 [g]_m=\int_{-\pi}^{\pi} \frac{dk}{2\pi} e^{-\ci k m}g(k), 
\end{equation}
so that the symbol can be reconstructed as
\begin{equation}
 g(k)=\sum_{m\in \mathbb{Z}} [g]_m e^{\ci k m}.
\end{equation}
We are concerned with semi-infinite matrices $T(g)$, with elements given by
\begin{equation}
 T(g)_{ij}=[g]_{i-j}
\end{equation}
for $i,j\in \mathbb{N}$. The word Toeplitz refers to the fact that the matrix elements only depend on $i-j$, while the notation $T(g)$ allows to keep track of the underlying symbol. Our aim is to compute the inverse of $T(g)$ exactly. Before proceeding, let us introduce a key concept, on which all the results below rely on: the Wiener-Hopf factorization. This is a decomposition
\begin{equation}
 g(k)=g^-(k)g^+(k)
\end{equation}
where $g^+$ (resp. $g^-$) only has nonnegative (resp. negative) Fourier coefficents, that is, $[g^+]_m=0$ for $m<0$, while $[g^-]_m=0$ for $m> 0$. Achieving a Wiener-Hopf factorization can be difficult in general, but it simply follows from
\begin{equation}
 g(k)=\exp\left(\sum_{m<0}[\log g]_m e^{\ci k m}\right)\exp\left(\sum_{m\geq 0}[\log g]_m e^{\ci k m}\right)
\end{equation}
in case $g$ has a well-defined logarithm. Our convention for the decomposition implies $[g^-]_0=1$. The main use of this decomposition is that it provides a simple way of writing an UL decomposition (similar to the famous LU decomposition) of the matrix, which makes computing the inverse much easier. Indeed $T(g^-)_{ij}=[g^-]_{i-j}=0$ if $i>j$ so $T(g^-)$ is upper triangular, while $T(g^+)$ is lower triangular. Their product gives
\begin{align}
 \left(T(g^-)T(g^+)\right)_{ij}&=\sum_{m=0}^\infty [g^-]_{i-m}[g^+]_{m-j}\\
 &=\sum_{m\in \mathbb{Z}} [g^-]_{i-m}[g^+]_{m-j}\\
 &=\int_{-\pi}^{\pi} \frac{dk}{2\pi}\int_{-\pi}^{\pi}\frac{dk'}{2\pi} e^{-\ci (ki-k'j)}g^-(k)g^+(k')\sum_{m\in\mathbb{Z}} e^{\ci m(k-k')}\\
 &=T(g)_{ij}
\end{align}
where we noticed that $[g^+]_{m-j}$ is always zero for $m<0$ to get the second line, and recognized a delta function in the third line. One can easily check by a similar calculation that the inverses of $U$ and $L$ are simply
\begin{equation}
 T(g^\pm)^{-1}=T(1/g^\pm).
\end{equation}
The inverse follows from the UL decomposition, since
\begin{equation}
 T(g)^{-1}=T(g^+)^{-1}T(g^-)^{-1},
\end{equation}
so
\begin{equation}\label{eq:inverse1}
 \left(T(g)^{-1}\right)_{ij}=\sum_{m=0}^{\infty}[1/g^+]_{i-m}[1/g^-]_{m-j}.
\end{equation}
The sum in the previous equation extends only to $m=\min(i,j)$, but it is convenient to keep this form. Let us now assume that $g$ is analytic in some complex neighborhood of the real axis. This means we can also write
\begin{equation}
 [1/g^-]_m=\int_{-\pi-\ci \eta}^{\pi-\ci \eta} \frac{dk'}{2\pi} e^{-\ci k' m}\frac{1}{g^-(k')}
\end{equation}
for some $\eta>0$, and the exact formula (\ref{eq:inverse1}) can be rewritten as
\begin{align}\label{eq:intergeom}
 \left(T(g)^{-1}\right)_{ij}&=\int_{-\pi}^{\pi} \frac{dk}{2\pi}\int_{-\pi-\ci \eta}^{\pi-\ci \eta} \frac{dk'}{2\pi} e^{-\ci (ki-k'j)} \frac{1}{g^+(k)g^-(k')}\sum_{m=0}^\infty e^{\ci m (k-k')}\\
 &=\int_{-\pi}^{\pi} \frac{dk}{2\pi}\int_{-\pi-\ci \eta}^{\pi-\ci \eta} \frac{dk'}{2\pi} e^{-\ci (ki-k'j)} \frac{1}{g^+(k)g^-(k')} \frac{1}{1-e^{\ci (k-k')}} \label{eq:inverse2}
\end{align}
where the point behind deforming the $k'$ integration contour was to make the geometric series in (\ref{eq:intergeom}) convergent. Equation (\ref{eq:inverse2}) will be crucial in the next appendix.
\section{Exact propagator}
\label{app:derivation}
\subsection{Wick's theorem}
In this part we use standard free fermions methods to express the general two point function as a ratio of determinant, which we then simplify. For convenience, we think of a fermion chain with sites in $\{1,\ldots,L\}$ first, and assume that $H$ is any quadratic Hamiltonian. The general propagator studied here is
\begin{equation}
 K_{ij}=\frac{\braket{\psi|e^{\tau_1 H}c_i^\dag e^{\tau_2 H}c_j e^{\tau_3 H}|\psi}}{\braket{\psi|e^{(\tau_1+\tau_2+\tau_3)H}|\psi}}
\end{equation}
for $\tau_1,\tau_2,\tau_3 \in \mathbb{C}$. We take $\ket{\psi}$ to be any real space product state of the form
\begin{equation}
 \ket{\psi}=c_{s(1)}^\dag \ldots c_{s(l)}^\dag \ket{0}
\end{equation}
for $\{s(1),\ldots,s(l)\}$ some ordered subset of $\{1,\ldots,L\}$. The propagator reads
\begin{equation}
 K_{ij}=\frac{\braket{0|c_{s(1)}\ldots c_{s(l)}  c_i^\dag(\tau_1)c_j(\tau_1+\tau_2)  c_{s(1)}^\dag(\tau_1+\tau_2+\tau_3)\ldots c_{s(l)}^\dag(\tau_1+\tau_2+\tau_3)|0}}
 {\braket{0|c_{s(1)}\ldots c_{s(l)}c_{s(1)}^\dag(\tau_1+\tau_2+\tau_3)\ldots c_{s(l)}^\dag(\tau_1+\tau_2+\tau_3)|0}},
\end{equation}
where we used the notation $c_s^\dag(\tau)=e^{\tau H} c_s^\dag e^{-\tau H}$. Since $H$ is quadratic (and conserves fermion number), $c_s^\dag(\tau)$ is a linear combination of the $c_{s'}^\dag$ only, and we can apply Wick's theorem both on the numerator and denominator. We get the ratio of determinants
\begin{equation}
 K_{ij}=\frac{\det\left(\begin{array}{cc}0&u\\v&M\end{array}\right)}
 {\det \left(\begin{array}{cc}1&0\\0&M\end{array}\right)},
\end{equation}
where $M$ is an $l\times l$ matrix with elements $M_{ab}=\braket{0|c_a c_b^\dag(\tau_1+\tau_2+\tau_3)|0}$, $u$ a $l$-line vector with elements $\braket{0|c_a c_i^\dag(\tau_1)|0}$ and $v$ a $l-$column vector with elements $\braket{0|c_j(\tau_1+\tau_2)c_b^\dag(\tau_1+\tau_2+\tau_3)|0}$. On the denominator, we have artificially enlarged the size of the matrix, so that dimensions on the numerator and denominator match.

Now, suppose we are able to invert the matrix $M$. By antilinearity of the determinant, we get
\begin{equation}
 K_{ij}=\sum_{a,b=1}^l \braket{0|c_a c_i^\dag (\tau_1)|0} (M^{-1})_{ab} \braket{0|c_j(\tau_1+\tau_2)c_b^\dag(\tau_1+\tau_2+\tau_3)|0}.
\end{equation}
Of course, inverting $M$ analytically is in general a hopeless task, but the formula can be useful numerically. In fact, we used it to generate some of the pictures shown in the paper. From this formula it is also possible to study infinite chains with an infinite number of particles, simply by sending $L\to \infty$ first, and then $l\to \infty$. To make progress, we need some extra assumptions on the form of the Hamiltonian, as well as the initial state. This is done in the next subsection.
\subsection{Contour integral formulas}
We now consider the limit $L\to \infty$, and assume that the Hamiltonian is invariant with respect to translations of one lattice site. In band theory language, the dispersion $\ve(k)$ is scalar, and we also assume $\ve(-k)=\ve(k)$. Translational invariance implies that the matrix element $M_{ab}=\int \frac{dk}{2\pi}e^{-\ci k(s(a)-s(b))}e^{(\tau_1+\tau_2+\tau_3)\varepsilon(k)}$ depends only on $s(a)-s(b)$.

For the class of initial states ($\ket{\psi_n}$) studied in the present paper, another simplification occurs, since $s(a)=na$, then $s(a)-s(b)=n(a-b)=s(a-b)$. Hence the matrix elements depend only on $a-b$, so we are dealing with a semi-infinite Toeplitz matrix. Then, the inverse simply follows from  formula (\ref{eq:inverse2}) in appendix.~\ref{app:toeplitz}.

The next step is to identify the correct symbol $g_n$ corresponding to $\ket{\psi_n}$. For $n=1$ this is obviously $g_1(k)=e^{(\tau_1+\tau_2+\tau_3)\varepsilon(k)}$. Writing the Fourier series $g_1(k)=\sum_{m\in \mathbb{Z}}[g_1]_m e^{\ci k m}$, we have by definition $g_n(k)=\sum_{m\in \mathbb{Z}}[g_1]_{mn}e^{\ci k m}$. Said differently, the $m-$th Fourier coefficient of $g_n$ is the $mn$-th Fourier coefficient of $g_1$. Inserting once again the integral representation, we get
\begin{equation}\label{eq:gdef}
 g_n(k)=\frac{1}{n}\sum_{p=0}^{n-1} e^{(\tau_1+\tau_2+\tau_3)\varepsilon(\frac{k+2p\pi}{n})}
\end{equation}
after a quick calculation. We now have all the necessary ingredients to compute the propagator. We have
\begin{align}
 K_{ij}&=\sum_{a,b=-\infty}^0 T(e^{\tau_1 \varepsilon})_{i,na}\left(T(g_n)^{-1}\right)_{ab} T(e^{\tau_3\varepsilon})_{nb,j}\\
 &=\sum_{a,b\in \mathbb{Z}} \int_{-\pi}^{\pi} \frac{dk}{2\pi}e^{-\ci k(i-na)}e^{\tau_1\varepsilon(k)}
 \int_{-\pi}^{\pi} \frac{dq}{2\pi}\int_{-\pi-\ci \eta}^{\pi-\ci \eta} \frac{dq'}{2\pi}\frac{e^{-\ci (qa-q'b)}}{g_n^+(q)g_n^-(q')}\frac{1}{1-e^{\ci (q-q')}}
 \int_{-\pi}^{\pi} \frac{dk'}{2\pi}e^{-\ci k'(nb-j)}e^{\tau_3\varepsilon(k')}\\
 &=\int_{-\pi}^{\pi} \frac{dk}{2\pi}\int_{-\pi+\ci \eta}^{\pi+\ci \eta} \frac{dk'}{2\pi} \frac{e^{-\ci (ki-k'j)}}{1-e^{-\ci n (k-k')}} \frac{e^{\tau_1\varepsilon(k)+\tau_3 \varepsilon(k')}}{g_n^-(nk)g_n^+(nk')}.
\end{align}
To get the last line, we recognized twice the series representation of a Dirac-delta distribution.

Now, we introduce the notation $g_n(k)=e^{(\tau_1+\tau_2+\tau_3)\varepsilon_n(k)}$, so that given (\ref{eq:gdef}), $\ve_n$ coincides with (\ref{eq:effective_disp}). The Wiener-Hopf decomposition $\varepsilon_n(k)=\varepsilon_n^+(k)+\varepsilon_n^-(k)$ provides us with a Wiener-Hopf factorization of $g_n$. In the following, we will also make use of the Hilbert transform $\ci\tilde{\varepsilon}_n(k)=\varepsilon_n^+(k)-\varepsilon_n^-(k)$. The above equation can be rewritten, with $\tau_1=R-y'$, $\tau_2=y'-y$, $\tau_3=R+y$, as 
\begin{equation}
 K_{ij}=\int_{-\pi}^{\pi} \frac{dk}{2\pi}\int_{-\pi+\ci \eta}^{\pi+\ci \eta} \frac{dk'}{2\pi} \frac{e^{-\ci (ki-k'j)}}{1-e^{-\ci n (k-k')}}
 e^{R(\ve(k)-\ve_n(nk)+\ve(k')-\ve_n(nk'))}e^{-y(\ve(k)-\ve(k'))+\ci R(\tve_n(nk)-\tve_n(nk'))}.
\end{equation}
Finally, the choice $\ve(k)=\cos k+\alpha \cos 2k$ gives formula (\ref{eq:main_exact}) advertised in the main text.

It is also possible to put the result in a slightly different form. Introducing the slightly more general notation $g_n(k|\tau)=\frac{1}{n}\sum_{p=0}^{n-1}e^{\tau \ve(\frac{k+2p\pi}{n})}$, and assuming for simplicity that $i,j$ are multiples of $n$, the propagator can be expressed as 
\begin{align}
 K_{ij}&=\sum_{a,b=-\infty}^0 T(g_n(.|\tau_1))_{i/n,a}\left(T(g_n(.|\tau_1+\tau_2+\tau_3))^{-1}\right)_{ab} T(g_n(.|\tau_3))_{b,j/n}\\
 &=\int_{-\pi}^{\pi} \frac{dk}{2\pi}\int_{-\pi+\ci \eta}^{\pi+\ci \eta} \frac{dk'}{2\pi} \frac{e^{-\ci (k\frac{i}{n}-k'\frac{j}{n})}}{1-e^{-\ci (k-k')}}
 \frac{g_n(k|R-y')g_n(k'|R+y)}{g_n^-(k|2R)g_n^+(k'|2R)}.
\end{align}
This formula can serve as an alternative starting point for the saddle point analysis, leading to the same results. 

Let us finally comment on issues related to the order of limit to establish the exact formula (\ref{eq:main_exact}). As mentioned in the introduction, there are possible ambiguities associated to the fact that the fermion chain is infinite. A proper regularization is to consider finite chains with sites $\in \{-L,-L+1,\ldots,L-1,L\}$ with (for example) open boundary conditions, compute the propagator for finite $L$, and only then take the $L\to \infty$ limit. Here we used formulas for the inverses of semi-infinite Toeplitz matrices, which already assume $L$ is infinite. However, there are also inversion formulas for large Toeplitz (or possibly Toeplitz+Hankel) matrices \cite{Widom1974}, which are very similar and would ultimately lead to the same result. To confirm this, we also checked the exact formula to high precision, by inverting the finite matrices using standard numerical linear algebra routines.
\bibliography{biblio}

\end{document}